\documentclass[11pt,a4paper]{article}

\usepackage[margin=1in]{geometry} 
\usepackage{amsmath} 
\usepackage{booktabs} 
\usepackage[authoryear]{natbib}

\usepackage{multirow}
\usepackage{graphicx,epstopdf}
\usepackage{subcaption} 
\usepackage{color} 
\usepackage{float}
\usepackage[titletoc,title]{appendix}

\usepackage{soul}

\usepackage{authblk}

\newcommand{\bmath}[1]{\boldsymbol{#1}}

\setcitestyle{authoryear,open={(},close={)}}

\begin{document}

\title{\Large A synthetic likelihood-based Laplace approximation for efficient design of biological processes}
\author{Dehideniya, M. B.$^{1}$, Overstall, A. M.$^2$, Drovandi, C. C.$^1$ and McGree, J. M.$^{*1}$ \\
~ \\
$^*$Corresponding author
 ~ \\
$^1$School of Mathematical Sciences \\
Queensland University of Technology \\
~ \\
$^2$Southampton Statistical Sciences Research Institute \\
University of Southampton}
\maketitle

\begin{abstract}

Complex models used  to describe biological processes in epidemiology and ecology  often have computationally intractable or expensive likelihoods.
This poses significant challenges in terms of Bayesian inference but more significantly in the design of experiments.
Bayesian designs are found by maximising the expectation of a utility function  over a design space, and typically this requires sampling from or approximating a large number of posterior distributions.
This renders approaches adopted in inference computationally infeasible to implement in design. 
Consequently, optimal design in such fields has been limited to a small number of dimensions or a restricted range of utility functions.
To overcome such limitations, we propose a synthetic likelihood-based Laplace approximation for approximating
 utility functions for models with intractable likelihoods.
As will be seen, the proposed approximation is flexible in that a wide range of utility functions can be considered, and remains computationally efficient in high dimensions.  To explore the validity of this approximation, an illustrative example from epidemiology is considered. 
Then, our approach is used  to design experiments with  a relatively large number of observations in two motivating applications from epidemiology and ecology.

\end{abstract}

~ \\
~ \\
\textbf{Keywords: }
Bayesian design, Foot and mouth disease, Kullback-Leibler divergence, Mutual information, Negative squared error loss, Prey and predator model, Total entropy.




\section{Introduction}
\label{s:intro}

Designing experiments to collect data that are as informative as possible about the process of interest is an important task in scientific investigation in, for instance, epidemiology \citep{Orsel2007},  system biology \citep{Faller2003} and ecology \citep{fenlon_faddy_2006}.
These biological systems require the development and use of realistic statistical models that involve computationally intensive or intractable likelihoods.
Unfortunately, this poses significant challenges in design of experiments, and has led to a number of recent developments, see \cite{Ryan2016} for a review.  

Common objectives in design of experiments include parameter estimation and model discrimination, and these have been considered for models with intractable likelihoods.  The Kullback Liebler distance (KLD) between the prior and the posterior has been considered by  \cite{Cook2008} as a utility function to find designs for parameter estimation.  Two stochastic models from epidemiology were considered, and the moment closure method was used to efficiently approximate the likelihood.
Such models have also been considered within an approximate Bayesian computation (ABC) framework where utilities were approximated based on summaries of the posterior distribution \citep{Drovandi2013,Price2016}.
In the presence of model uncertainty, such methods were extended by \cite{Dehideniya2017} to find designs for discriminating between competing intractable likelihood models where the ABC rejection algorithm for model choice \citep{Grelaud2009} was used to estimate posterior quantities. 
Extensions to dual purpose experiments for model discrimination and parameter estimation have also been proposed by \cite{Dehideniya2018TESL} where the total entropy utility \citep{Borth1975} was considered. 
To approximate this utility, a synthetic likelihood approach for discrete data was developed, and this allowed a wide variety of utility functions to be considered in design for models with intractable likelihoods.  However, further developments are needed to extend this approach to realistically-sized experiments.

Until recent work by \cite{Overstall2018IIL}, designs for models with intractable likelihoods have been limited to a small number of  design dimensions.
\cite{Overstall2018IIL} used emulation within an indirect inference framework to avoid the evaluation of computationally expensive or intractable likelihoods, and were able to consider design spaces of an order of magnitude larger than what had been considered previously.
However, their approach is limited to  likelihood-based utility functions, that is, utility functions that can be expressed in terms of the likelihood and/or the marginal likelihood.
In this paper, we address this limitation by extending the work of \cite{Dehideniya2018TESL} to higher dimensions thus allowing a wide variety of utility functions to be considered when designing large-scale experiments for models with intractable likelihoods. 
In our approach, we use summary statistics to avoid the curse of dimensionality, and also develop a Laplace-based approximation to the posterior distribution. 
This enables fast utility approximation, and thus allows designs to be found in reasonable time frames.

This paper is outlined as follows. 
In Section \ref{s:background}, we provide details of the types of models considered in this paper and the methods currently used for estimating such models. 
In Section \ref{s:BayExPDesign}, Bayesian experimental design is described along with the utility functions considered in this paper.
Section \ref{s:infFramework} presents the proposed synthetic likelihood-based Laplace approximation for models with intractable likelihoods.
Then, an illustrative example is considered in Section \ref{s:example} along with two motivating examples for this work.  
Finally, a discussion of our work is provided in Section 6 along with some  limitations and suggestions for future research directions.  


\section{Bayesian inference for intractable likelihood models}
\label{s:background}

Applications considered in this paper focus on biological processes observed at $L$ time points $\bmath{d} = \{t_1,t_2, \ldots , t_L\}$ such that $t_{i+1} > t_{i}; i = 1,2,...,L-1$. 
The observed data $\bmath{y} = \{\bmath{y}_1,\bmath{y}_2, \ldots, \bmath{y}_L\}$ from such experiments can be described by a continuous time Markov chain (CTMC) model where each data point is the number of individuals in particular states of the Markov process. 
Through the Markov property, the likelihood of $\bmath{y}$ is given by:

\begin{equation}
\label{likelihood_y}
p(\bmath{y}|\bmath{\theta}, \bmath{d}) = \prod_{i=0}^{L}  p(y_{i+1}|y_i,\bmath{\theta}), 
\end{equation} 

where $p(y_{i+1}|y_i,\bmath{\theta})$ is the probability that the observed state of the process is $y_{i+1}$ at time $t_{i+1}$ given that at time $t_{i}$ the observed state of the process is $y_{i}$, and $\bmath{\theta}$ is a $q$-vector of model parameters.
Note that $y_0$ in Equation (\ref{likelihood_y}) is the initial, known condition of the process at time $t = 0$.
Within a Bayesian framework, prior knowledge about $\bmath{\theta}$ is  represented by the prior distribution $p(\bmath{\theta})$ and is updated by the observed data  $\bmath{y}$ via the likelihood function to obtain the posterior distribution $p(\bmath{\theta} | \bmath{y}, \bmath{d})$ using Bayes' theorem.  That is:

\[
p(\bmath{\theta} | \bmath{y}, \bmath{d}) = \frac{ p(\bmath{\theta}) p(\bmath{y}|\bmath{\theta}, \bmath{d})}{p(\bmath{y}|\bmath{d})},
\]

where $p(\bmath{y}|\bmath{d}) = \int_{\bmath{\theta}} p(\bmath{\theta}) p(\bmath{y}|\bmath{\theta}, \bmath{d}) \mbox{d}\bmath{\theta}$ is known as the model evidence.

For Markov processes with a large number of states, transition probabilities  $p(y_{i+1}|y_i,\bmath{\theta})$  cannot be evaluated analytically and consequently, the likelihood function is not available in closed-form.  
Thus, likelihood-free methods are generally used for making inference about such models.

Likelihood-free methods such as ABC rejection  \citep{Beaumont2002} are based on the simulation of prior predictive data.
Then, these simulated data are compared with the observed data, and the posterior distribution of parameters is formed by collecting the parameter values that resulted in similar simulated data to that which were observed, where `similar' is measured by a suitably defined discrepancy function and pre-defined tolerance value.
As the number of observed data points increases, the chance of simulating a data set that is similar to what was observed decreases, reducing the computational efficiency of the algorithm.
To address this, a lower-dimensional set of summary statistics can be considered in place of the data.
However, despite this, in practice, ABC rejection suffers from low acceptance rates meaning that a large number of data sets need to be generated in order to obtain a reasonable approximation to the posterior distribution. 
%
%

Alternative approaches to likelihood-free inference include forming approximate parametric distributions for the summary statistics and  forming variational approximations to the posterior distribution. 
The synthetic likelihood approach \citep{WoodSL2010} forms an approximation to the likelihood by assuming the summary statistics follow a multivariate normal distribution.  \cite{Price2018} consider this within a Bayesian framework, and use Monte Carlo Markov chain (MCMC) to obtain samples from an approximate posterior distribution.
Alternatively, variational approximations \citep{Bishop2006} can be considered where the posterior is approximated by a parametric probability distribution $q_{\lambda}(\bmath{\theta})$ where $\lambda$ is the variational parameter.
In forming this approximation, $\lambda$ is optimised to minimise the  Kullback-Leibler  divergence  between $q_{\lambda}(\bmath{\theta})$ and $p(\bmath{\theta} | \bmath{y}, \bmath{d})$. 
\cite{Tran2017} proposed a variational approximation for models with intractable likelihoods, and  \cite{Ong2018} considered this within a synthetic likelihood approach to replace computationally expensive likelihood evaluations.
In both approaches, the variational parameter $\lambda$ is optimised using a stochastic gradient method to obtain the approximation of the posterior distribution of parameters.


When there are $K$ competing models to describe the process of interest, each with prior model probability $p(m)$ and model parameters $\bmath{\theta}_m$ with prior distributions $p(\bmath{\theta}_m|m)$, model selection is performed based on the posterior model probability given by:


\begin{equation}
\label{eq:SLLA_pmp}
p(m | \bmath{y} ,\bmath{d}) = \frac{p(\bmath{y}| m ,\bmath{d})\, p(m)}{\sum_{m=1}^K p(\bmath{y}| m ,\bmath{d})\, p(m)},
\end{equation}

where $p(\bmath{y}| m ,\bmath{d}) = \int_{\bmath{\theta}_m} p(\bmath{y}|\bmath{\theta}_m, m ,\bmath{d}) p(\bmath{\theta}_m|m)\mbox{d}\bmath{\theta}_m$ and $p(\bmath{y}|\bmath{\theta}_m, m ,\bmath{d})$ is the likelihood function for model $m$.

However, as can be seen, evaluating the posterior model probability relies on being able to evaluate the likelihood.  When this is unavailable, the ABC rejection algorithm for model choice \citep{Grelaud2009} can be used.  Here, data are simulated conditional on both the model and model parameters, and compared with the observed data. 
The model that generates similar data sets most frequently is selected as the most appropriate model.
Alternatively, approximations to the likelihood can be used.  These include synthetic likelihood-based approximations and variational approaches.


%
%
%
%
%



\section{Bayesian experimental designs}
\label{s:BayExPDesign}

%
%

A properly planned experiment will provide informative  data for subsequent inferences of interest, such as parameter estimation, model discrimination and/or prediction.
Thus, prior to any experimentation, the values of all controllable variables need to be carefully chosen to ensure the data are as informative as possible.
In the context of experimental design, a set of possible values for these controllable variables is  defined  as the design,
and the informativeness of data obtained from such a design is measured by a utility function which is defined according to the experimental goal/s.
In this paper, we focus on when to observe a biological process of interest, thus time is the design/controllable variable.
The selection of the values for the controllable variables can be defined as a decision problem under the uncertainty about the parameters $ \bmath{\theta}_m$, model $m$ and data $\bmath{y}$ yet to be observed under design $\bmath{d}$ \citep{Lindley1978}.
Therefore, the expectation of the utility function (across all such uncertainty) is considered in finding optimal designs.
This expected utility can be defined as follows:  

\begin{equation}\label{e:ExpUtyMD}
U(\bmath{d}) = \sum_{m=1}^{K} p(m)\left\{\int_{\bmath{y}} \int_{\bmath{\theta}_m} u(\bmath{d},\bmath{y}, \bmath{\theta}_m , m)\, p(\bmath{y} \, | \, \bmath{\theta}_m, \,m,\bmath{d}) \, p(\bmath{\theta}_m|m) \, d\bmath{\theta}_m \,d\bmath{y} \right\},
\end{equation}

where the utility $u(\bmath{d},\bmath{y}, \bmath{\theta}_m , m)$ is specified to encapsulate the purpose of the experiment such as estimation of parameters across all the $K$ competing models \citep{McGree2016}, discriminating between competing models \citep{Drovandi2014} or dual goals of parameter estimation and model discrimination \citep{Borth1975,McGree2017}.
When the utility function $u(.)$ is independent from the model parameter  $\bmath{\theta}_m$, Equation (\ref{e:ExpUtyMD}) can be re-expressed as follows: 

\begin{equation}\label{e:ExpUtyMDWithoutPara}
U(\bmath{d}) = \sum_{m=1}^{K} p(m)\left\{\int_{\bmath{y}} u(\bmath{d},\bmath{y},m)\, p(\bmath{y} \, | \,m,\bmath{d})\,d\bmath{y} \right\}.
\end{equation}

Unfortunately, the expected utility is generally not available in closed-form and thus needs to be approximated.  The most common approach adopted in the literature for this is Monte Carlo integration.
When  $u(.)$ depends on the model parameters  $\bmath{\theta}_m$, the approximate expected utility of design $\bmath{d}$ can be expressed as follows:

\begin{equation}
\label{e:Est_MC_ExpUty_1}
\hat{U}(\bmath{d}) =   \sum_{m=1}^{K} p(m) \frac{1}{Q} \sum_{j =1}^{Q}  {u}(\bmath{d},\bmath{y}_m^j,\bmath{\theta}_m^j,m),
\end{equation}

where $(\bmath{y}^j_m,\bmath{\theta}^j_m)$ are generated from the joint distribution of $(\bmath{y},\bmath{\theta}_m)$ given model $m$ and design $\bmath{d}$.  It is here where we assume that, despite the likelihood being intractable, it is straightforward to simulate data from the model.  We note that such an assumption is typical when dealing with intractable likelihoods models, for example, see \cite{Beaumont2002}.

Similarly, when $u(.)$ is independent from  $\bmath{\theta}_m$, the expected utility given by Equation (\ref{e:ExpUtyMDWithoutPara}) can be approximated as,

\begin{equation}
\label{e:Est_MC_ExpUty_2}
\hat{U}(\bmath{d}) =   \sum_{m=1}^{K} p(m) \frac{1}{Q} \sum_{j =1}^{Q}  {u}(\bmath{d},\bmath{y}_m^j,m),
\end{equation}

where $\bmath{y}_m^j$ is a data set generated from model $m$ under design $\bmath{d}$. 


The accuracy of the above Monte Carlo approximation to the expected utility increases as the number of Monte Carlo samples $Q$ increases for each model $m$ but so does the computational burden. 
\cite{Drovandi2018} demonstrated that the accuracy of $\hat{U}(\bmath{d})$ for a given number of Monte Carlo samples can be increased by using randomised Quasi-Monte Carlo (RQMC) methods.
Following \cite{Drovandi2018}, here a Sobol sequence $(0,1]^{(q_m+n)}$ is used to first generate $q_m$ parameters $\bmath{\theta}_m$ of model $m$ and then simulate $n$ observations $\bmath{y}$ from the model conditional on  $\bmath{\theta}_m$ and $\bmath{d}$.
In order to obtain an unbiased estimate of ${U}(\bmath{d})$, these deterministic sequences are randomised by using the Owen type \citep{Owen1997} of scrambling implemented in \cite{randtoolbox}  with different seed values each time a sequence is generated. 
In our implementation, system time was used as the seed value for each simulated sequence.



\subsection{Utility functions for parameter estimation}

In this work, we consider two utility functions for designing efficient experiments to estimate parameters of models with intractable likelihoods, namely the Shannon information gain on the parameters  and the negative squared error loss utility. These are defined below.

\subsubsection{Shannon information gain on $\bmath{\theta}_m$}

The Shannon information gain on the parameters  $\bmath{\theta}_m$ (SIGP) has been commonly used as a utility function to design experiments for estimating parameters, for instance,  \cite{Overstall2017}. 
The SIGP utility is given by: 

\begin{equation} \label{UtyParKL}
u_{\mathrm{SIGP}}(\bmath{d},\bmath{y},\bmath{\theta}_m,m) = \log p(\bmath{\theta}_m|m) - \log p(\bmath{\theta}_m|\bmath{y},m,\bmath{d}).
\end{equation}

Maximising the expectation of this utility is equivalent to maximising the expected KLD between the prior and posterior distribution of $\bmath{\theta}_m$, see \cite{Ryan2003}.

\subsubsection{Negative squared error loss utility}

When the goal of the experiment is to obtain point estimates of model parameters, the negative squared error loss (NSEL) utility can be used. 
Given that $\bmath{\theta}_m$ is a vector of $q_m$ elements, the NSEL utility is:

\begin{equation} \label{UtyNSEL}
u_{\mathrm{NSEL}}(\bmath{d},\bmath{y},\bmath{\theta}_m,m) = -\sum_{i=1}^{q_m} \big(  \theta_{im} - E[ {\theta}_{im}|\bmath{y},\bmath{d},m] \big)^2,
\end{equation}
where $\theta_{im}$ is the $i^{th}$ parameter value of $\bmath{\theta}_m$. 

\subsection{Shannon information gain on the model indicator}

Model discrimination utilities are considered when the goal of the experiment is to determine the most appropriate model for the data.  For this purpose, the Shannon information gain on the model indicator (SIGM) has been considered to design experiments \citep{Overstall2018NAB}.  Such a utility can be expressed as follows:

\begin{equation} \label{UtyMDMI}
u_{\mathrm{SIGM}}(\bmath{d},\bmath{y},m) = \log p(m)  -  \log p(m \, | \,\bmath{y},\bmath{d}).
\end{equation}

Maximising the expectation of this utility is equivalent to maximising the expected mutual information between the model indicator and the data \citep{Box1967,Drovandi2014}.  


\subsection{Shannon information gain on $\bmath{\theta}_m$ and the model indicator} 

Experimental goals of parameter estimation and model discrimination can be considered simultaneously by formulating dual-purpose utility functions.  One such formulation was proposed by \cite{Borth1975} who considered the total entropy about the model and model parameters.  This is equivalent to considering the Shannon information gain on the joint distribution of $\bmath{\theta}_m$ and the model indicator (SIGT), and thus can be expressed as follows: 

\begin{equation}\label{eq:TE_utility}
u_{\mathrm{SIGT}}(\bmath{d},\bmath{y},m) = \log p(\bmath{\theta}_m,m) - \log p(\bmath{\theta}_m,m|\bmath{y},\bmath{d}).
\end{equation}


\bigskip
In each case, the utility function depends on the analytically intractable posterior distribution. Therefore, we need to approximate the posterior distribution for each evaluation of the utility. Given that the approximate utility is embedded within a Monte Carlo approximation to the expected utility, we will need a total of $K \times Q$ approximations to the posterior distribution for just one evaluation of the Monte Carlo approximation to the expected utility. Then the approximate expected utility needs to be maximised over the design space. This is a computationally demanding task even for tractable likelihood models. In order facilitate efficient design for intractable likelihood models, fast methods for approximating posteriors are needed.
In the next section, we describe our proposed synthetic likelihood-based Laplace approximation for this purpose, and then demonstrate its use in an illustrative and two motivating examples.

\section{Synthetic likelihood-based Laplace approximation}
\label{s:infFramework}

Our proposed method for forming computationally efficient approximations to posterior distributions for models with intractable likelihoods is based on the synthetic likelihood in combination with the Laplace approach for approximate posterior inference. To describe this method, we start by defining the synthetic likelihood, and then describe the Laplace approximation.

The synthetic likelihood approach is a method of approximating the likelihood of observed data $\bmath{y} = \{\bmath{y}_1,\bmath{y}_2, \ldots, \bmath{y}_L\}$ given $\bmath{\theta}_m$ for intractable likelihood model $m$ \citep{WoodSL2010}. 
This is achieved by assuming that the summary statistics for model $m$, $\bmath{S}_m$, conditional on $\bmath{\theta}_m$, follow a multivariate normal distribution with mean $\bmath{\mu}(\bmath{\theta}_m)$ and variance-covariance $\Sigma(\bmath{\theta}_m)$.
In general, $\bmath{\mu}(\bmath{\theta}_m)$ and $\Sigma(\bmath{\theta}_m)$ cannot be evaluated analytically for a given value of $\bmath{\theta}_m$ but can be approximated by simulating $n$ data sets from the model conditional on $\bmath{\theta}_m$ and evaluating the summary statistics for each data set.
This yields a distribution of summary statistics for which $\bmath{\mu}(\bmath{\theta}_m)$ and $\Sigma(\bmath{\theta}_m)$ can be estimated. 
Then, the log-likelihood of observed summary statistics $\bmath{s}_{obs}$ under model $m$ can be approximated as follows:

\begin{equation} 
l_{s}(\bmath{s}_{obs}|\bmath{\theta}_m,m,\bmath{d}) = - \frac{1}{2} \Big[ \log |\hat{\Sigma}(\bmath{\theta}_m)| - (\bmath{s}_{obs}- \hat{{\bmath{\mu}}}(\bmath{\theta}_m))^T \, \hat{\Sigma}(\bmath{\theta}_m) (\bmath{s}_{obs}- \hat{{\bmath{\mu}}}(\bmath{\theta}_m)) + L \log(2\pi)   \Big], 
\end{equation}
where  $\hat{\bmath{\mu}}(\bmath{\theta}_m)$ and $\hat{\Sigma}(\bmath{\theta}_m)  $ are the estimated mean vector and variance-covariance matrix of the simulated summary statistics from the model of interest with parameter $\bmath{\theta}_m$. 


Given the above approximation to the log-likelihood, a posterior distribution can be found.  However, for design, efficiency is key, and accordingly, previous studies have considered importance sampling for models with tractable \citep
{Weir2007,McGree2012} and intractable \citep{Dehideniya2018TESL} likelihoods as a fast approximation when the distance between the prior and posterior is relatively small.  
However, as the number of observations increases, the resultant posterior distribution can be considerably different from the prior, and importance sampling 
may provide an inefficient approximation to the posterior. 
One alternative that has been used in Bayesian design is the Laplace approximation \citep{Long2013,Overstall2018NAB}.
Suppose we have observed data $\bmath{y}$ under design $\bmath{d}$ generated from model $m$ with parameters $\bmath{\theta}_m$. 
Then, the Laplace approximation is found by first locating the mode of the posterior distribution of $\bmath{\theta}_m$ based on the synthetic likelihood as follows:

\[
\bmath{\theta}^*_m = \max_{\bmath{\theta}_m}~ l_s(s_{obs}|\bmath{\theta}_m,m,\bmath{d}) + \log p(\bmath{\theta}_m|m),
\]

then evaluating the Hessian at this point $H(\bmath{\theta}^*_m)$.  The approximation to the posterior is a multivariate normal distribution of the following form:

\begin{equation}\label{eq:phat}
    \hat{p}(\bmath{\theta}_m | \bmath{y}, m, \bmath{d}) = \mbox{MVN}(\bmath{\theta}^*_m, \Sigma^*_m),
\end{equation}

where $\Sigma^*_m = [-H(\bmath{\theta}^*_m)]^{-1}$.


Thus, this approximation to the posterior distribution simply requires maximising the posterior density (through the choice of $\bmath{\theta}_m$), and evaluating the Hessian at this mode.  As pointed out by \cite{WoodSL2010}, due to small-scale noise associated with evaluating   $l_{s}(\bmath{s}_{obs}|\bmath{\theta}_m,m,\bmath{d})$,  derivative-based, numerical optimisation approaches cannot be used to find the posterior mode. 
Thus, in this work the Nelder-Mead algorithm for derivative-free optimisation  \citep{Kelley1999} is used  to find  the parameter value which maximises  the posterior density.
To approximate the Hessian matrix, methods proposed by \cite{Fasiolo2016PHDTh} can be considered where the Hessian of the synthetic likelihood function is estimated for a given model $m$ at a parameter value $\bmath{\theta}_m$ based on a set of local regression models fitted between model parameters and summary statistics (see Algorithm 2 on page 61 of \cite{Fasiolo2016PHDTh}).

Once this approximation has been formed, it can be substituted into the expressions for the utility functions given in the previous section such that they can be approximated efficiently.  That is, for SIGP: 

\[
\hat{u}_{\mathrm{SIGP}}(\bmath{d},\bmath{y},\bmath{\theta}_m,m) = \log p(\bmath{\theta}_m|m) - \log \hat{p}(\bmath{\theta}_m|\bmath{y},m,\bmath{d}),
\]

where $\hat{p}(\bmath{\theta}_m | \bmath{y}, m, \bmath{d})$ is given in Equation (\ref{eq:phat}).




Following   \cite{Overstall2018NAB}, the NSEL utility given in Equation (\ref{UtyNSEL}) can be evaluated efficiently, just requiring the posterior mode to be found.  That is, $E[ {\theta}_{im}|\bmath{y},m,\bmath{d}] \approx {\theta}^*_{im}$.

For evaluating the discrimination utility, a computationally efficient approximation to the model evidence is needed.  This can be obtained via the Laplace approximation as follows: 

\begin{equation}
\label{eq:SLLA_evi}
\hat{p}(\bmath{y}|m, \bmath{d}) =  (2\pi)^{\frac{q_m}{2}} |H(\bmath{\theta}_m^*)^{-1}| \, \, p(\bmath{y}|   \bmath{\theta}_m^*,m,\bmath{d}) \,\, p(\bmath{\theta}_m^*|m),
\end{equation}

where the synthetic likelihood is used to approximate the likelihood.

Thus, the discrimination utility in Equation (\ref{UtyMDMI}) can be approximated as follows:

\[
\hat{u}_{\mathrm{SIGM}}(\bmath{d},\bmath{y},m) = \log p(m)  -  \log \hat{p}(m | ,\bmath{y},\bmath{d}),
\]

where $\hat{p}(m | ,\bmath{y},\bmath{d})$ is evaluated as shown in Equation (\ref{eq:SLLA_pmp}) based on the above approximation to the model evidence. 

Both of the above approximate utilities can then be used to approximate the SIGT utility given in Equation (\ref{eq:TE_utility}).  This approximation has the following form:

\[
\hat{u}_{\mathrm{SIGT}}(\bmath{d},\bmath{y},\bmath{\theta}_m,m) = \log p(\bmath{\theta}_m|m)-\log \hat{p}(\bmath{\theta}_m|\bmath{y},m,\bmath{d}) + \log p(m) -  \log \hat{p}(m | ,\bmath{y},\bmath{d}).
\]


Adopting the above approach for Bayesian inference in design has a number of advantages over the recent  work of \cite{Dehideniya2018TESL} who proposed a synthetic likelihood approximation using the full data set with continuity correction for discrete observations.
Although their approximation was shown to work well when a few data points were observed,  it becomes computationally expensive and inefficient as the number of observations increases. 
In contrast, the synthetic likelihood approach based on summary statistics provides a computationally feasible method of approximating the likelihood of a large number of observations. 
Thus, the proposed approach is more suitable for designing experiments which yield a large number of observations, where large is defined in the context of design for models with intractable likelihoods.

Another advantage of our proposed approach is that the Laplace approximation requires less likelihood evaluations when compared to alternative posterior approximations such as importance sampling and Markov chain Monte Carlo.
Consequently, the use of Laplace approximation reduces the number of data sets that need to be simulated from the model.
While model simulation is generally an efficient process, having to repeat this a large number of times imposes significant computational burden.
Indeed, in the Bayesian experimental design literature, pre-simulated data have been used to avoid the computational cost of simulating a large number of data sets during the optimisation, for instance,  \cite{Drovandi2013,Hainy2016}.
While this may provide some computational efficiency, this results in having to consider a  discrete design space, and therefore potentially suboptimal designs will be found.
In adopting our proposed methods, a continuous design space can be considered which should lead to the location of designs that perform better with respect to the experimental goal/s.

%

\section{Examples}
\label{s:example}

In this section, we consider one illustrative example and two motivating examples to demonstrate the benefits and practical implications of our proposed methodologies.
First, the proposed utility approximation is validated by evaluating the SIGT utility of designs for  dual purpose experiments for two epidemiological models namely the death model and Susceptible-Infected (SI) model. 
Secondly, the performance of our approach is demonstrated through designing experiments to learn about foot and mouth disease. 
Finally, as the third example, we consider    designing laboratory microcosm experiments to estimate parameters of a prey and predictor model found in ecology.

Locating optimal designs involves maximising $\hat{U}(\bmath{d})$ over a design space.   
In this work, we used the approximate coordinate exchange (ACE) algorithm \citep{Overstall2017} to find optimal designs in a continuous design space. 
The ACE algorithm iteratively optimises one design variable at a time by emulating the expected utility of the given design dimension, and optimising the predicted value as given by the emulator.
At each iteration, the newly found design is compared with the current design, and is selected based on a Bayesian hypothesis test.
When implementing ACE, a number of tuning parameters need to be specified. 
In the examples described in the following section, 5000 and 500 Monte Carlo samples were used for the hypothesis test and constructing the emulator, respectively.
Otherwise, the default settings for ACE were used as given in the R-package \texttt{acebayes} \citep{Overstall2017ACE_RPaper}.

\subsection{Example 1 - Dual purpose designs for the death and SI models}

The death and SI models  can be used to describe the spread of a disease among a closed population of size $N$.
In this example, optimal time points which yield informative observations for both discriminating between these competing models and estimating parameters of the models are considered.

The death model \citep{Cook2008} divides the population into two sub-populations, susceptible and infected. 
The state of the CTMC at time $t$ is defined as the number of infected individuals at time $t$, ${I(t)}$. 
Given $I(t)=i$, the transition probability of the possible state at $t+\Delta_t $ is given by 
$$ P\big[\, i + 1 \, | \, i \, \big] = {\beta}_1\,(N-i)\,\Delta_t + \mathcal{O}(\Delta_t ),$$

where ${\beta}_1$ is the rate at which susceptible individuals become infected.

The SI model \citep{Cook2008} assumes that the infected individuals in the population also contribute to the spread of diseases, represented by an additional parameter ${\beta}_2$. 
Given that $I(t)=i$, the transition probability of the possible state at $t+\Delta_t $ is given by

$$ P \big[ \, i + 1 \, | \,i \, \big] = ({\beta}_1 + {\beta}_2\, i)\,(N-i)\,\Delta_t + \mathcal{O}(\Delta_t ).$$

The same priors for the unknown parameters considered by \cite{Dehideniya2018TESL} are used here, and they are as follows: 
$ \log(\beta_1) \sim N(-0.48,0.3^2)$ for Death model and $ \log(\beta_1) \sim  N(-1.1, 0.4^2)$ and $ \log(\beta_2) \sim N(-4.5, 0.63^2))$ for the SI model.  The prior predictive distribution of data from each model is shown in Figure 2a.  

In the context of experimental design for models with intractable likelihoods, forming informative summary statistics is a difficult task as the summaries need to be informative over the entire prior predictive distribution and the design space.
To handle this, we propose to use summary statistics that are informative across a subset of the design space, where this subset is defined by the perceived informativeness of a given design {\it a priori}.
That is, it seems reasonable that, in order to estimate the probability of becoming infectious given an individual is susceptible, data on individuals in both states is needed. 
It is this intuition that is used in subsetting the design space, and this is achieved through inspection of the prior predictive data from a given design. 
In terms of summary statistics, here we propose to use the mean and variance of the observed counts as these were shown to be informative across the prior predictive distribution for a random selection of designs (see Appendix \ref{ISS_DMSI}).

The Hessian approximation proposed in \cite{Fasiolo2016PHDTh} is based on a set of regression models fitted between the model parameters and summary statistics.
They also propose an additional  regression step, where each model parameter is regressed against the summary statistics, and the fitted model parameters are used as summary statistics to improve the scalability of the Hessian approximation as the number of summary statistics increases (see Section 4.5.1 of \cite{Fasiolo2016PHDTh}). 
In order to ensure reasonable accuracy in the approximation,  the validity of these additional linear regression models was assessed based on a measure of goodness-of-fit (coefficient of multiple determination, $R^2$).
This enables the identification of data sets and designs that yield poor approximations of the Hessian matrix and consequently the utility. 
Then, based on a defined threshold value for $R^2$, we substituted poor estimates of the utility function with a minimum utility value.
As such, the estimate of the expected utility will be down weighted, and potentially avoided within the optimisation. 
For applications considered in this paper, it was found that for models with a single parameter, a threshold value of around 0.7 could be used while for other models a lower threshold can be considered (around 0.1).
Obviously, when this occurs, this will introduce a bias in the estimation of the expected utility.  
Thus, we investigated the effect of this bias by comparing the expected utility of randomly selected designs evaluated based on the actual likelihood  to our synthetic likelihood approach, and these results are shown in Figure 1. 
	
\begin{figure}[ht]
		
		\centering
		\begin{subfigure}[b]{0.25\textwidth}
			\centering
			\includegraphics[width=2in,width=2in]{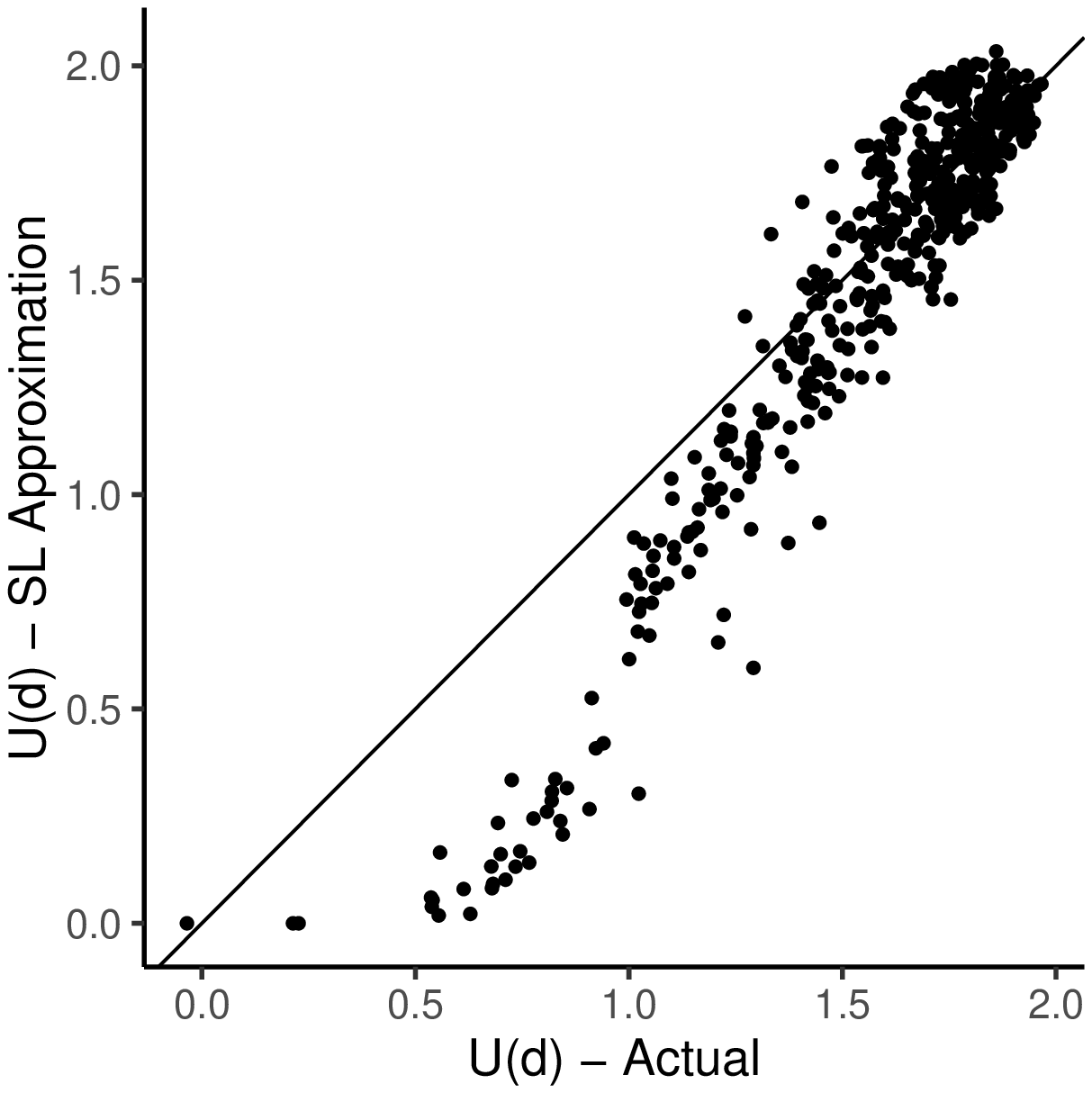}
			\caption{SIGT}
			\label{fig:Ex1TE_SynL}
		\end{subfigure}
		\hfill
		\begin{subfigure}[b]{0.25\textwidth}
			\centering
			\includegraphics[width=2in,width=2in]{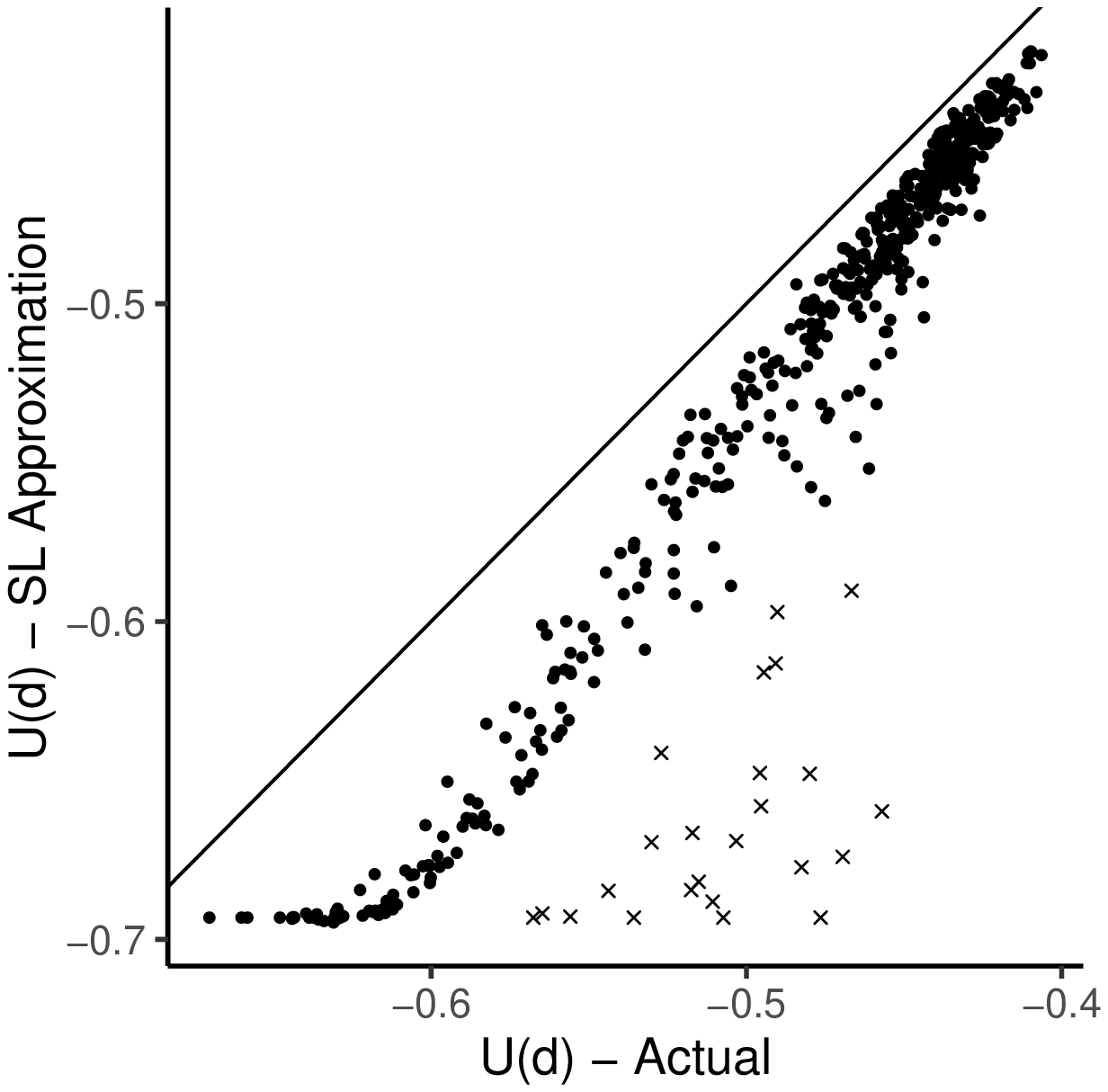}
			\caption{SIGM}
			\label{fig:Ex1MI_SynL}
		\end{subfigure}
		\hfill
		\begin{subfigure}[b]{0.25\textwidth}
			\centering
			\includegraphics[width=2in,width=2in]{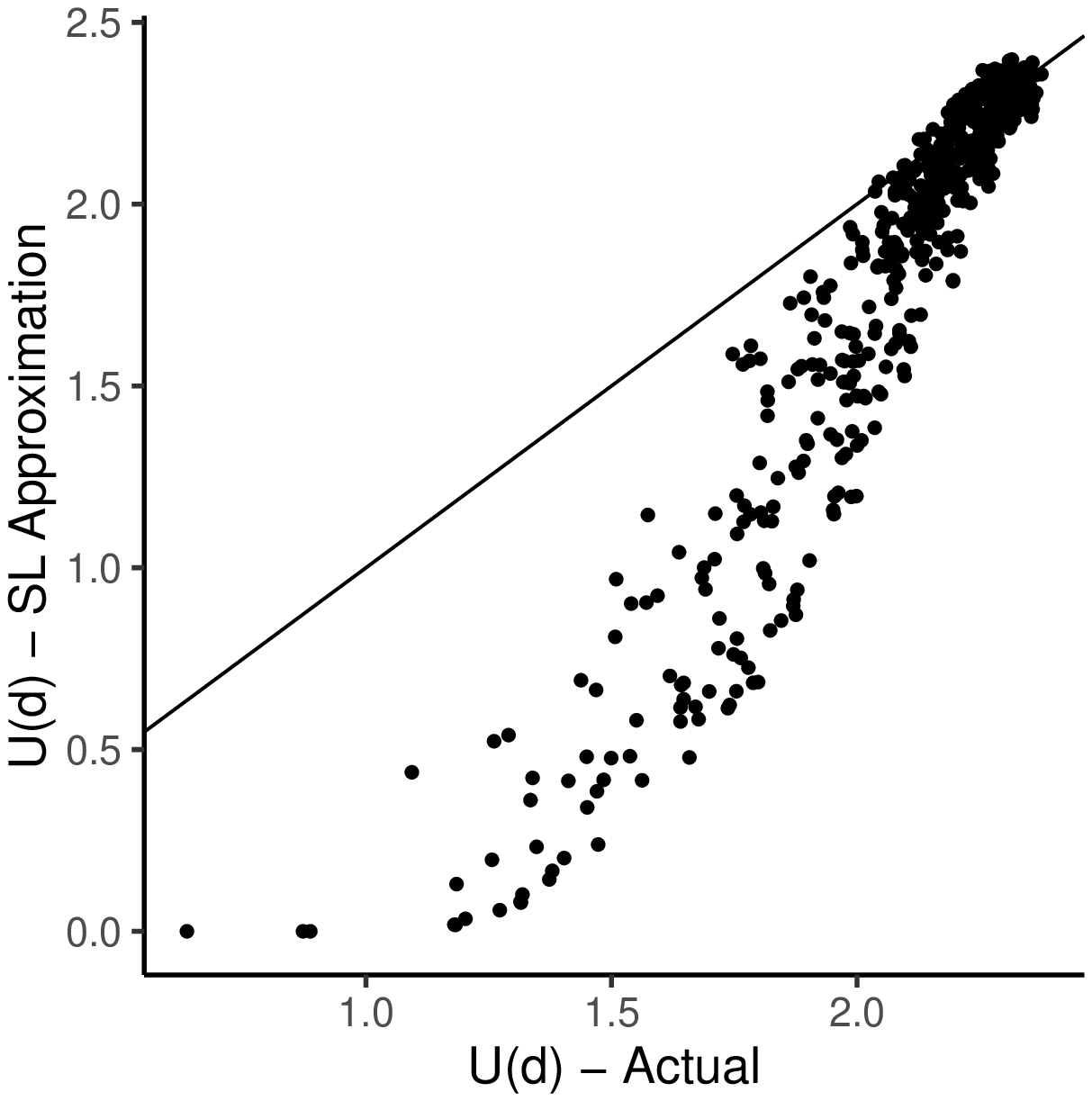}
			\caption{SIGP}
			\label{fig:Ex1KLD_SynL}
		\end{subfigure}
		\hfill

		\caption{Comparison of the estimated expected utility of design with 15 design points according to the (a) SIGT, (b) SIGM  and  (c) SIGP utilities  using Laplace approximation based on synthetic and actual likelihoods. Here, designs with biased estimate of the expected utility are represented by ($\times$). In each plot, $y=x$ line indicates a perfect match of approximated and actual utility evaluations.
		}
		\label{fig:ValEx1_SynL_ACT_DM_SI} 
	\end{figure}

As is evident from Figure \ref{fig:ValEx1_SynL_ACT_DM_SI}, the proposed approach preserves a monotonic relationship between the approximated and actual utility values, and reasonably approximates the utility values for designs with higher utility values.
It is noted that, the proposed approach provides a biased estimate of the SIGM utility for some designs, see ($\times$) in Figure \ref{fig:Ex1MI_SynL}.  
Given that  these designs have relatively low expected utility values under the actual utility evaluation, the maximisation of the expected utility should not be affected  by the introduced bias  in handling the poor approximation of the Hessian matrix.
	Therefore, we propose that our approximation can be used  to locate optimal designs.

\begin{figure}[H]
		\centering
		\centering
		\begin{subfigure}[b]{0.95\textwidth}
			\centering
			\includegraphics[width=5in,height=3in, keepaspectratio=false]{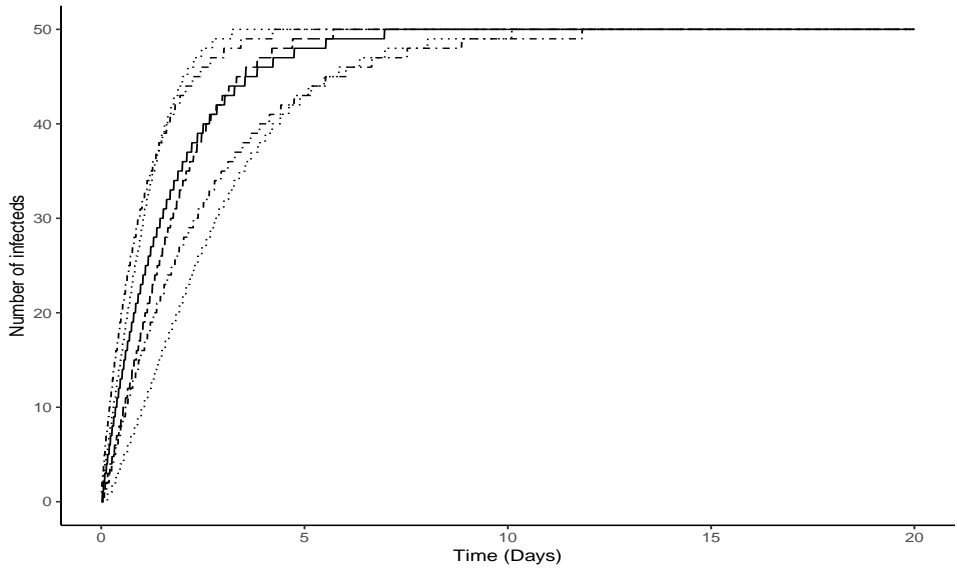}
			\caption{}
			\label{fig:Ex1PriorPred_DMSI_I}
		\end{subfigure}
		\hfill
		\centering
		\begin{subfigure}[b]{0.95\textwidth}
			\centering
			\includegraphics[ width=5.75in,height=3.5in, keepaspectratio=false]{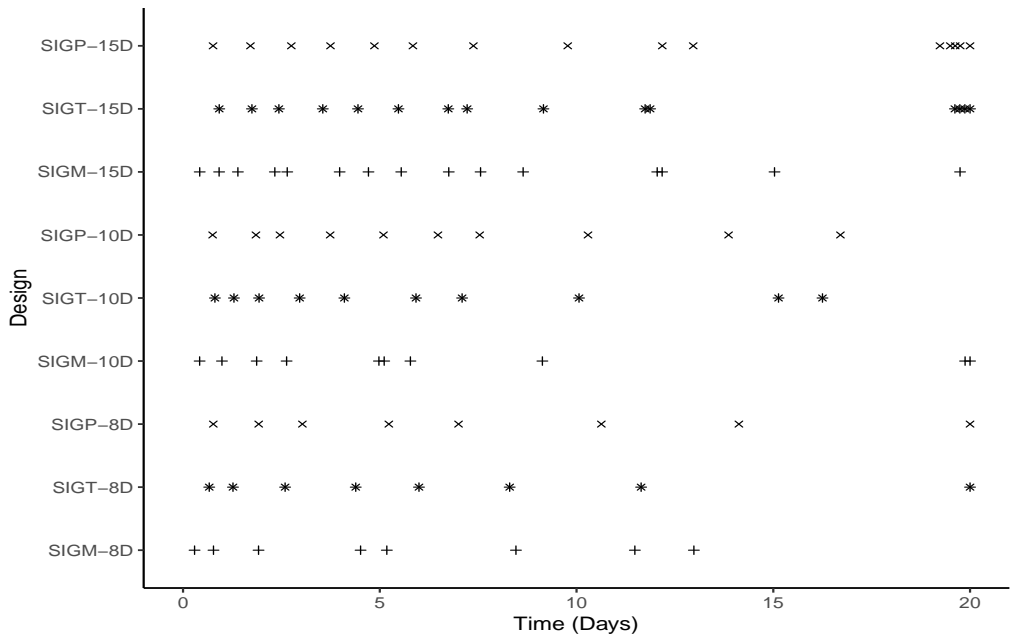}
			\caption{}
			\label{fig:OD_Ex1}
		\end{subfigure}
		\hfill
		\caption{(a) Prior predictive distribution of number of infecteds based on the death model (solid) and SI model (dashed). Here, dot-dashed and dotted lines represent the $10\%$ aand $90\%$ prior predictive quantiles of the death and SI model, respectively. (b) Optimal designs found under the SIGT utility ($*$) along with the SIGP ($\times$) and SIGM ($+$) utilities for the death and SI models.   }
		
	\end{figure}

	Optimal designs  under the SIGT, SIGM and SIGP utility functions were located by the ACE algorithm, and are shown in Figure \ref{fig:OD_Ex1}. 
	The expected utility values of the optimal designs were re-evaluated 100 times with different draws from the prior predictive distribution, and the mean and standard deviation of these expected utility values are given in Table \ref{table:Ex1_OD_TE_MI_KLD}.

	\begin{table}[h]
		\caption{Expected utility values (standard deviation) of optimal designs derived under different utility   functions.} \label{table:Ex1_OD_TE_MI_KLD}
		{\begin{tabular*}{\columnwidth}{@{}l@{\extracolsep{1cm}}l@{\extracolsep{1cm}}r@{}}
				\hline Utility function  & Number of  design points $|\bmath{d}|$ & $U(\bmath{d}^*)$ (SD) 
				\\
				\hline
				\multirow{3}{*} {SIGM } & 8 &  -0.444 (0.002) \\
				
				& 10 &   -0.434 (0.002)\\ 
				
				& 15 &  -0.423 (0.002) \\ 

				\hline
				\multirow{3}{*} {SIGT}  & 8 &  1.891 (0.006) \\ 
				
				& 10 &   1.931 (0.006)\\ 
				
				& 15 &  2.007 (0.007) \\ 

				\hline
				\multirow{3}{*} {SIGP}  & 8 & 2.328 (0.007) \\ 
				
				& 10 & 2.375 (0.007)\\ 
				
				& 15 &   2.433 (0.007) \\ 
				
				\hline
		\end{tabular*}}
		\bigskip
	\end{table}

\begin{figure}[ht]
		\centering
		\begin{subfigure}[b]{0.49\textwidth}
			\centering
			\includegraphics[width=3.4in,width=3in]{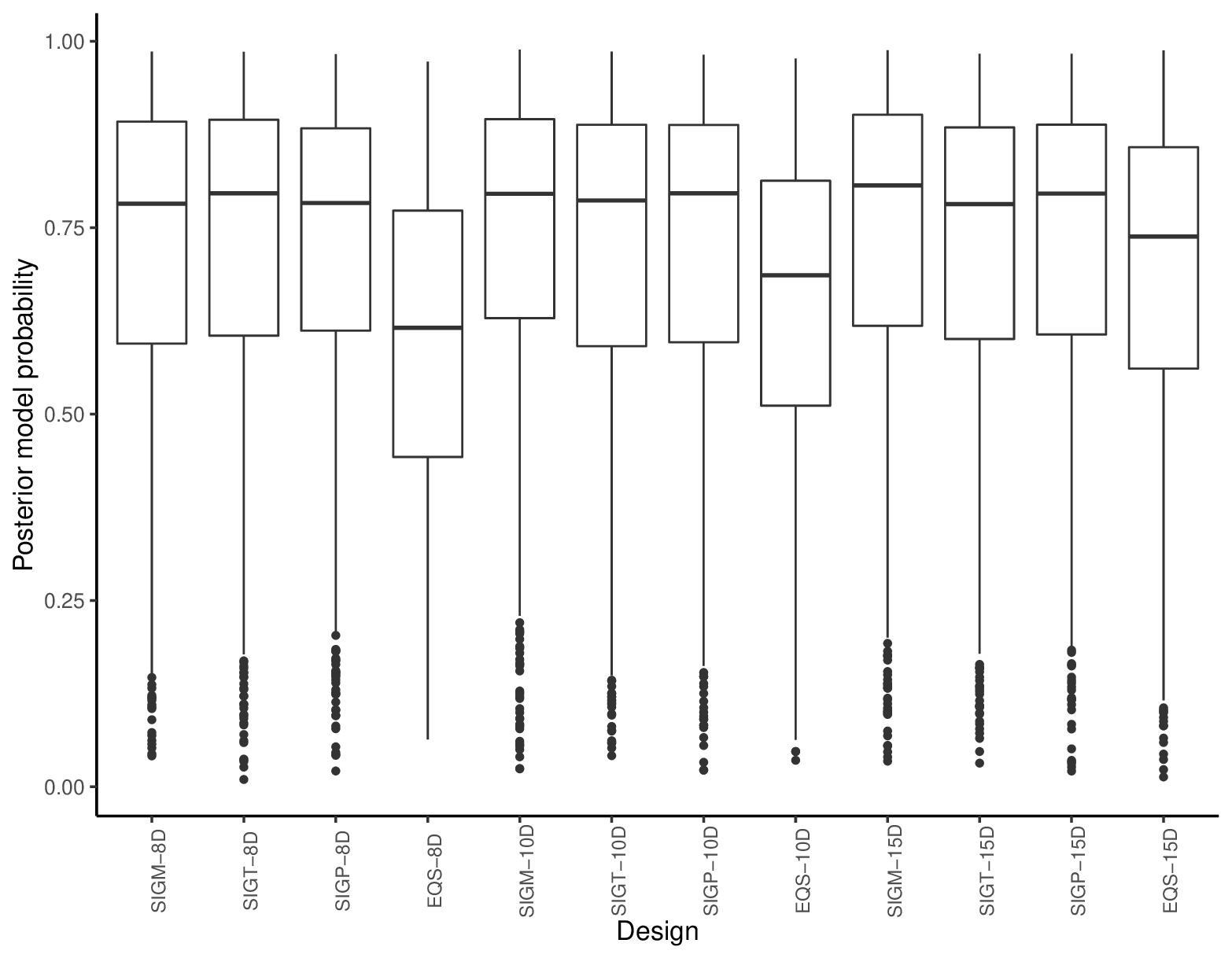}
			\caption{Death model}
			\label{fig_Ex1DM_MD}
		\end{subfigure}
		\hfill
		\begin{subfigure}[b]{0.49\textwidth}
			\centering
			\includegraphics[width=3.4in,width=3in]{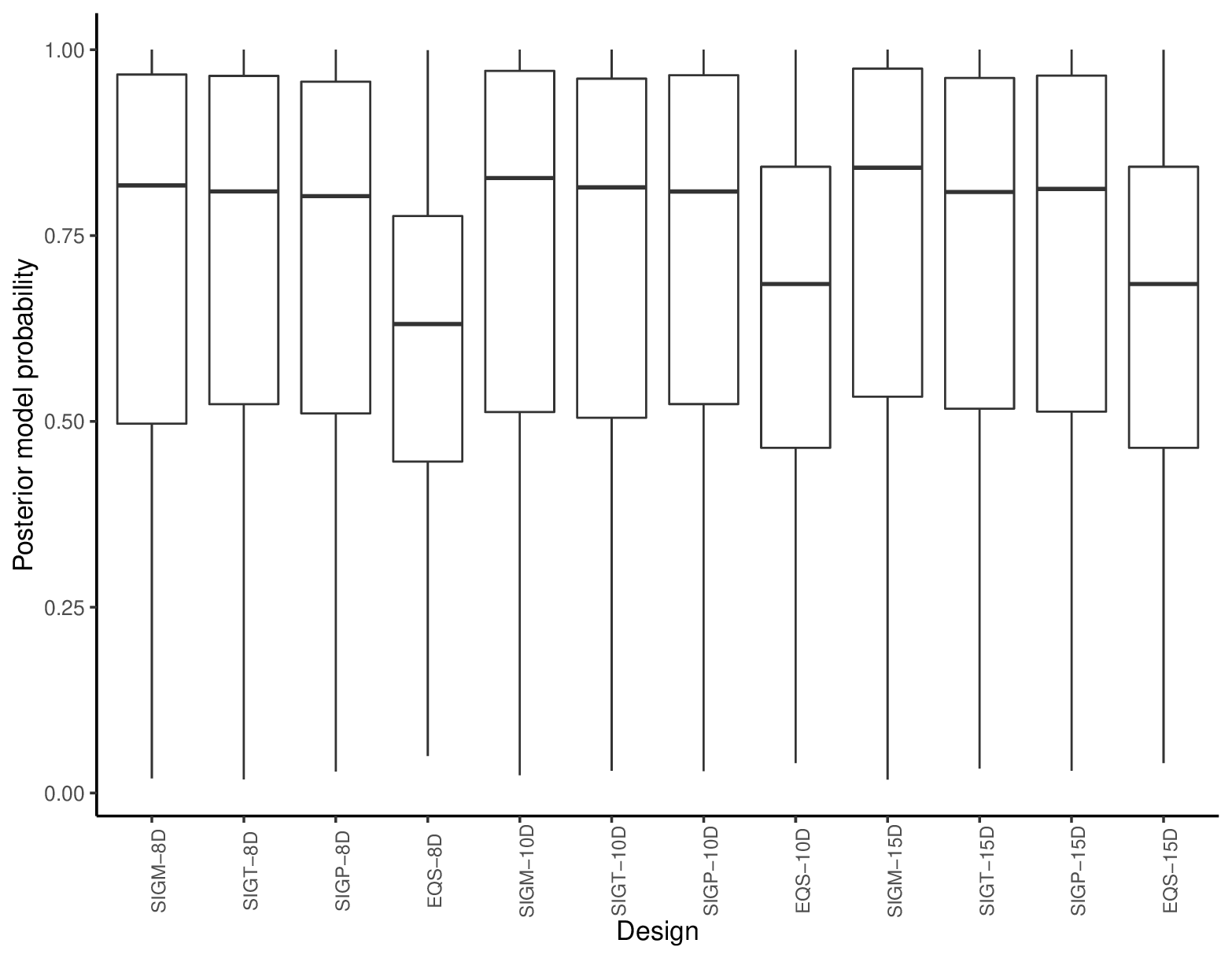}
			\caption{SI model}
			\label{fig_Ex1SI_MD}
		\end{subfigure}
		\caption{The posterior model probability of  the data generating model  obtained for observations generated from  (a) death model and (b) SI model according to optimal designs and an equally spaced designs.} 
		\label{fig:pmps_ActualEx1DM_SI} 
	\end{figure}

	The optimal designs based on SIGT were compared with the SIGM and SIGP designs in terms of addressing each experimental goal individually.
	Further, for each optimal design, an equally spaced design with same number of design points was also considered.
	In this comparison, for each design, 1000 data sets were simulated from both models, and posterior  inference was undertaken based on actual likelihood for the death model and approximated likelihood for SI model \citep{Sidje1998}.
	First, posterior model probabilities of the data generating model were estimated based on Laplace approximation as described by Equations (\ref{eq:SLLA_pmp}) and (\ref{eq:SLLA_evi}).
	These results are shown in Figure \ref{fig:pmps_ActualEx1DM_SI} where it is evident that all optimal designs perform equally well for model discrimination, with some advantage over the equally spaced design.

\begin{figure}[ht]
		\centering
		\begin{subfigure}[b]{0.49\textwidth}
			\centering
			\includegraphics[width=3.4in,width=3in]{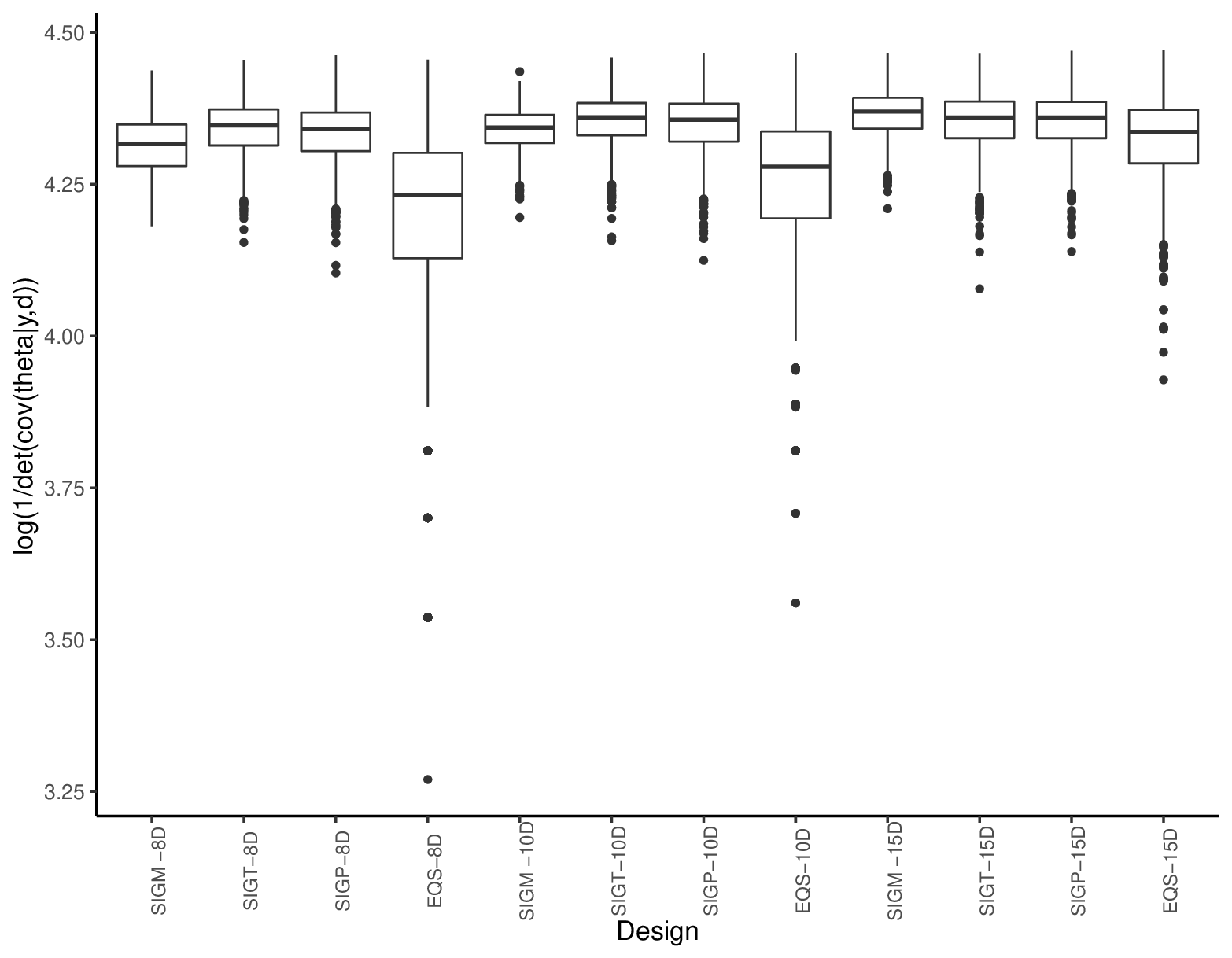}
			\caption{Death model}
			\label{fig_Ex1DM_PE}
		\end{subfigure}
		\hfill
		\begin{subfigure}[b]{0.49\textwidth}
			\centering
			\includegraphics[width=3.4in,width=3in]{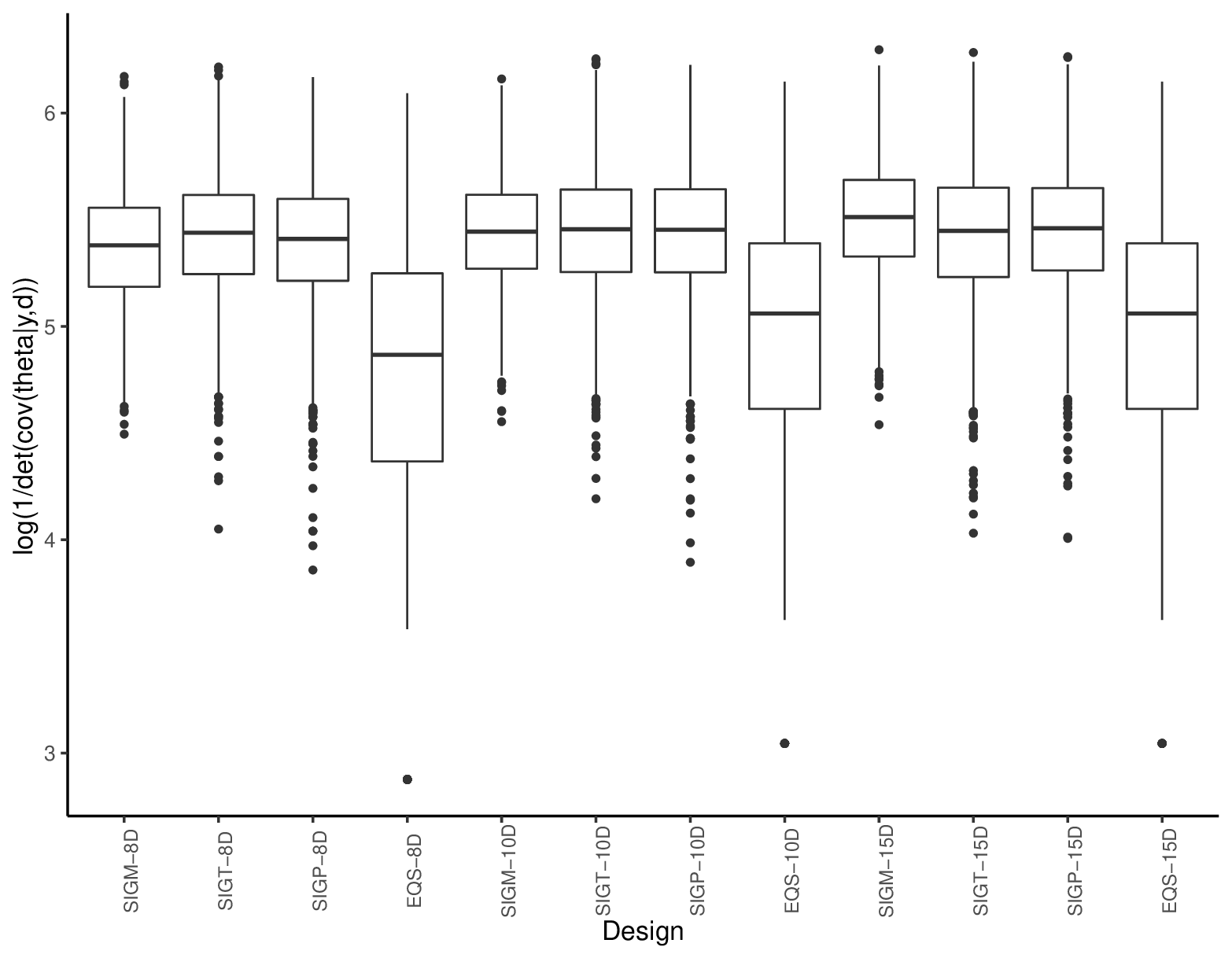}
			\caption{SI model}
			\label{fig_Ex1SI_PE}
		\end{subfigure}
		\caption{The log determinant of the inverse of posterior variance-covariance matrix of parameters of the data generating model when observations generated from  (a) death model and (b) SI model according to optimal designs and an equally spaced designs.} 
		\label{fig:LDPCV_ActualEx1DM_SI} 
	\end{figure}

	Secondly, the designs were compared in terms of parameter estimation. 
	For the death model, the log reciprocal of the posterior variance of $\beta_1$ and, for SI model, the log determinant of the inverse of posterior variance-covariance matrix of $(\beta_1, \beta_2)$ were used to measure the performance of the designs for parameter estimation. 
	In this validation step, Laplace importance sampling (LIS) was considered to more accurately approximate the posterior of the model parameters.
	LIS is a combination of the Laplace approximation and the importance sampling where the importance distribution is chosen based on the Laplace expansion, see \cite{kuk1999} for further details. 
	The results from this validation are shown in Figure \ref{fig:LDPCV_ActualEx1DM_SI} where it can be seen that all optimal designs perform similarly well, and consistently outperform the equally spaced design.

	\subsection{Example 2 -  Dual purpose designs for foot and month disease} 

	Foot and mouth disease (FMD) is a contagious disease which affects livestock such as cattle, pigs and sheep  \citep{Knight-Jones2013}.
	Both the Susceptible-Infectious-Recovered (SIR) model \citep{Orsel2007} and the Susceptible-Exposed-Infectious-Recovered (SEIR) model \citep{Backer2012} have been proposed to describe epidemic data collected from studying the spread of FMD over time.

	The SIR model assumes that susceptible individuals become infectious immediately after they make an infected-contact with the infectious individuals in the population.
	Then, the infectious individuals recover after time $T \sim \exp(\alpha)$  becoming immune and no longer spread the disease to other individuals.
	As in the previous example, the spread of FMD among $N$ individuals can be described by a CTMC model. 
	Given that there are $s$ susceptible and $i$ infectious individuals at time $t$,
	the probabilities of possible events in the next infinitesimal time period $\Delta_t$ are given by, 
	
	$$ P\big[s - 1 \, , i + 1 \, | \, s \, , i \big] = \frac{\beta \, s\,i}{N} \,\Delta_t + \mathcal{O}(\Delta_t ),$$
	
	$$P\big[s , i - 1 |s , i \big] =  \alpha \,\Delta_t + \mathcal{O}(\Delta_t ),$$
	where $\beta$  is the rate at which an infectious individual make infected-contacts per unit time. 
	
	In contrast, the SEIR model assumes that the susceptible individuals do not become infectious immediately after they been exposed to the disease, but after time $T_E \sim \exp(\alpha_E)$. 
	During this period, exposed individuals do not show any symptoms of being infected, and therefore the number of exposed individuals $e$ at time $t$ is unobservable. 
	Once the exposed individuals become infectious, they contribute to the spread of the disease and recover after time $T_I \sim \exp(\alpha_I)$. 
	Given that there are $s$ susceptible, $e$ exposed and $i$ infectious individuals at time $t$, the probabilities of possible events in the next infinitesimal time period $\Delta_t$ are given by,
	
	$$ P\big[s - 1 \, , e + 1 \, , i  \, | \, s \, ,e \, , i \big] = \frac{\beta \, s\,i}{N} \,\Delta_t + \mathcal{O}(\Delta_t ),$$  
	
	$$P\big[s \, , e - 1 \, , i +1  \, | \, s \, ,e \, , i\big] =  \alpha_E e\,\Delta_t + \mathcal{O}(\Delta_t ),$$
	
	$$P\big[s \, , e  \, , i- 1  \, | \, s \, ,e \, , i\big] =  \alpha_I i \,\Delta_t + \mathcal{O}(\Delta_t ),$$
	where $\beta$  is the rate at which an infectious individual make infected-contacts per unit time.

\begin{figure}[H]
		\begin{subfigure}[b]{1\textwidth}
			\centering
			\includegraphics[width=5in,height=2in]{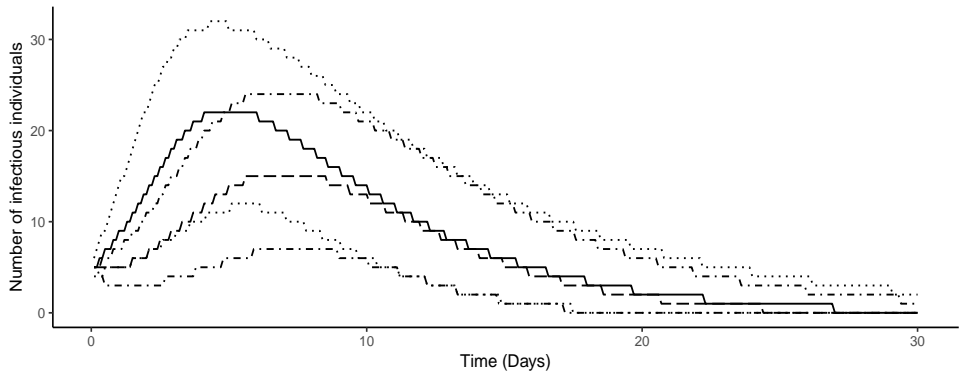}
			\caption{}
			\label{fig:P3Ex2SEIR_I}
		\end{subfigure}
		\hfill
		\begin{subfigure}[b]{1\textwidth}
			\centering
			\includegraphics[width=5in,height=2.3in]{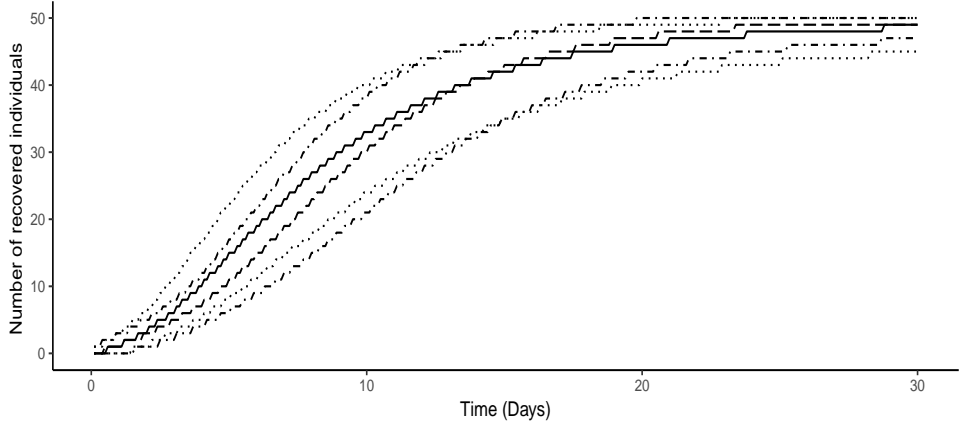}
			\caption{}
			\label{fig:P3Ex2SEIR_R}
		\end{subfigure}
		\hfill
		\begin{subfigure}[b]{1\textwidth}
			\centering
			\includegraphics[width=6in,height=3in, keepaspectratio=false]{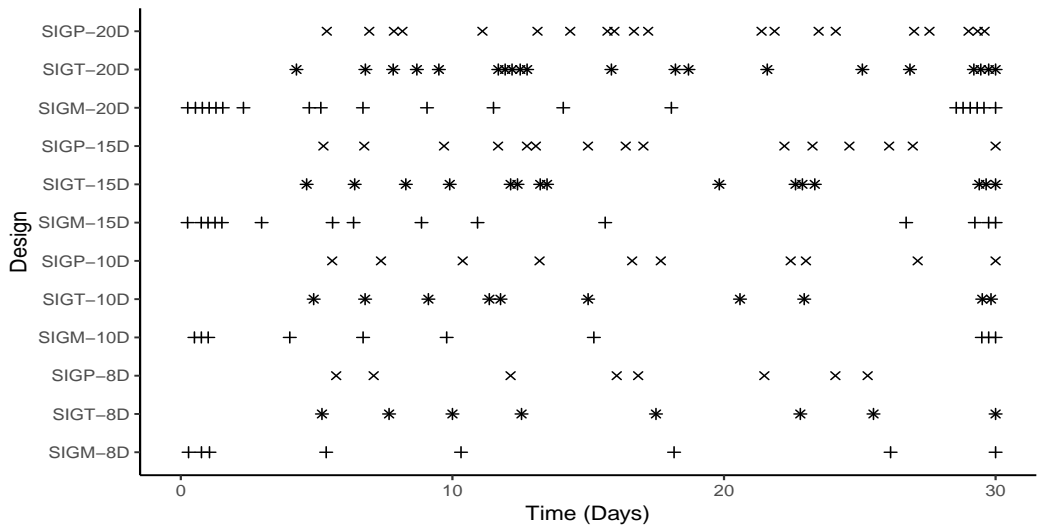}
			\caption{}
			\label{fig:P3Ex3LV2_OD}
		\end{subfigure}
		
		\caption{Prior predictive distributions of (a) infectious and (b) recovered individuals based on the SIR (solid) and SEIR (dashed) models. In both figures, dotted and dot-dashed lines represent the $10\%$ and $90\%$ quantiles of the prior predictive distributions, respectively. (c) Optimal designs found under SIGT utility ($*$) along with the SIGP ($\times$) and SIGM ($+$) utilities for the SIR and SEIR models. }
		\label{fig:OD_Ex2}       
	\end{figure}

	Following \cite{Dehideniya2018TESL}, $\log(\beta) \sim N(-0.09,0.19^2) $  and  $\log(\alpha) \sim N(-1.63,0.32^2) $ were chosen to describe the uncertainty about the parameters of the SIR model, and for SEIR model  $\log(\beta) \sim N( 0.44,0.16^2) $ , $\log(\alpha_E) \sim N(-0.69,0.2^2)$ and $\log(\alpha_I) \sim N(-1.31,0.38^2)$ were chosen as the priors, see plots of the prior predictive distributions given in Figures \ref{fig:P3Ex2SEIR_I} and \ref{fig:P3Ex2SEIR_R}.
	At the beginning of the experiment, $t=0$, there are 5 infectious and 45 susceptible    individuals and the population is observed starting  from $t=0.25$ days (6 hours) until 30 days.
	In order to obtain an observation schedule which is feasible to implement, these observation times were selected such that they are at least 0.25 days apart, and we consider up to 20 design points.
	At each observation time, the number of infectious $(I)$ and recovered $(R)$ individuals are recorded.
	In approximating the expected utility of designs,  mean, median and variance were considered as the summary statistics to approximate the synthetic likelihood of observed data as described in Section \ref{s:infFramework}.
	Optimal designs found under three utility functions are illustrated in Figure \ref{fig:OD_Ex2}c.
	The expected utility of each optimal design was re-evaluated 100 times and the mean and the standard deviation of those estimated  utilities are given in Table  \ref{table:Ex2_OD_TE_MI_KLD}.

	\begin{table}[h]
		\caption{ Expected utility values (standard deviation)  of optimal designs derived under different utility functions.} \label{table:Ex2_OD_TE_MI_KLD}
		{\begin{tabular*}{\columnwidth}{@{}l@{\extracolsep{1cm}}l@{\extracolsep{1cm}}r@{}}
				\hline Utility function  & Number of  design points $|\bmath{d}|$ & $U(\bmath{d}^*)$ (SD) 
				\\
				\hline
				\multirow{4}{*} {SIGM } & 8 &  -0.275  (0.005) \\
				
				& 10 &    -0.266  (0.005)\\ 
				
				& 15 &   {-0.267} (0.005) \\ 
				
				& 20 &   {-0.261} (0.005) \\ 
				
				\hline
				\multirow{4}{*} {SIGT}  & 8 &    1.152 (0.010) \\ 
				
				& 10 &    1.172 (0.008)\\ 
				
				& 15 &   1.196 (0.009) \\ 
				
				& 20 &    1.205 (0.009) \\ 
				
				\hline
				\multirow{4}{*} {SIGP}  & 8 & 1.569 (0.013) \\ 
				
				& 10 &  1.583 (0.008)   \\ 
				
				& 15 &   1.592 (0.009)     \\ 
				
				& 20 &  1.611 (0.007)   \\ 
				
				\hline
		\end{tabular*}}
		\bigskip
	\end{table}
	
\begin{figure}[ht]
		\centering
		\begin{subfigure}[b]{0.49\textwidth}
			\centering
			\includegraphics[width=3.4in,width=3in]{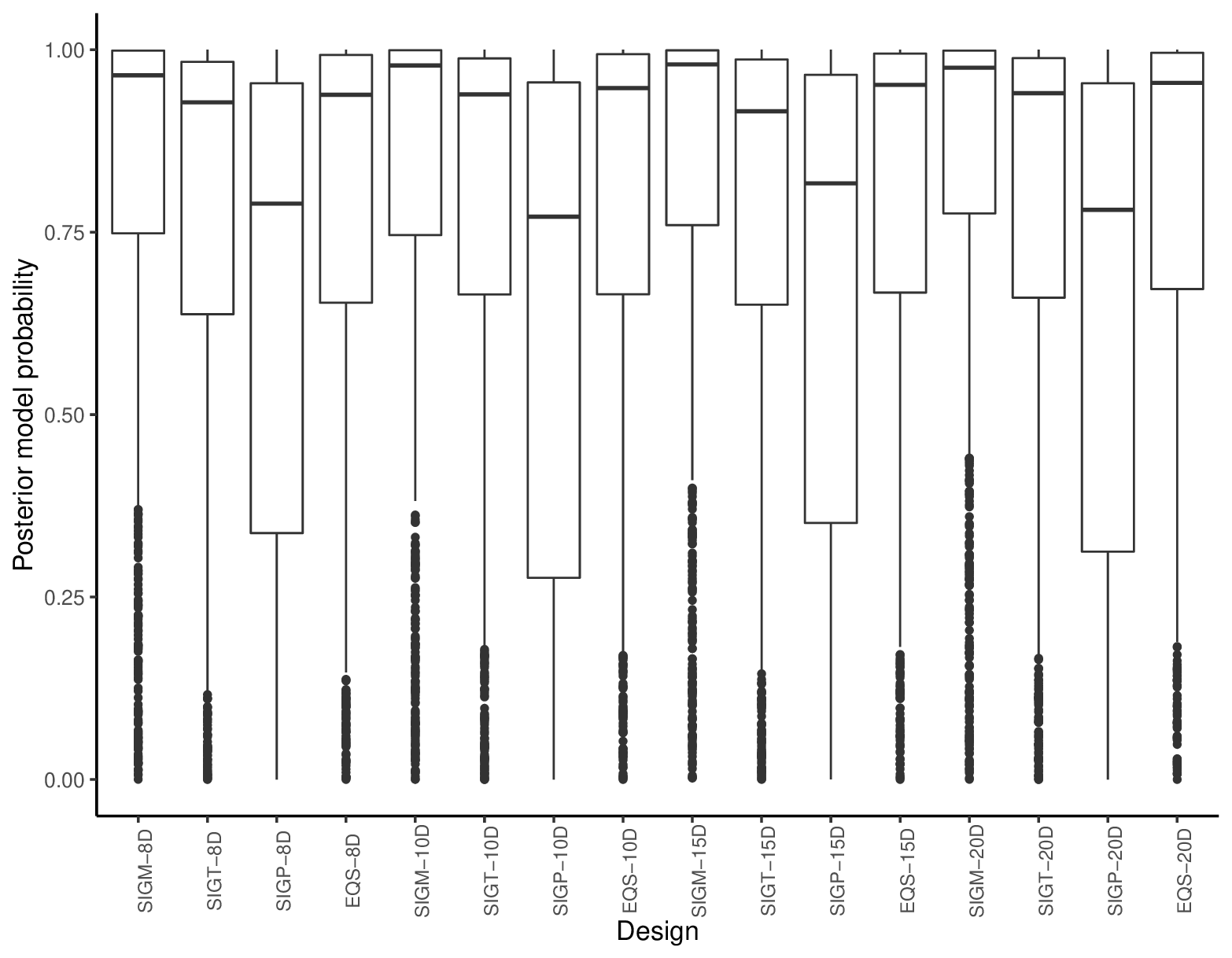}
			\caption{SIR model}
			\label{fig_Ex1SIR_MD}
		\end{subfigure}
		\hfill
		\begin{subfigure}[b]{0.49\textwidth}
			\centering
			\includegraphics[width=3.4in,width=3in]{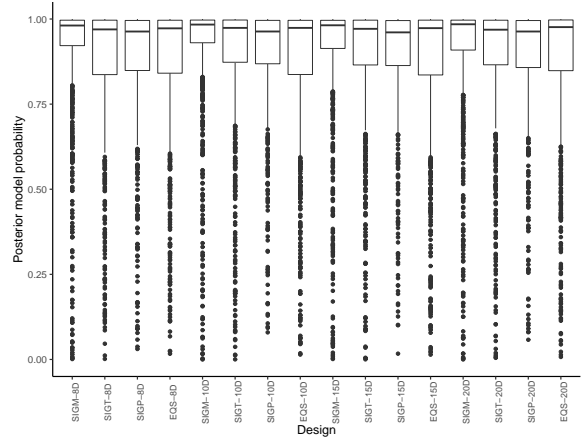}
			\caption{SEIR model}
			\label{fig_Ex2SEIR_MD}
		\end{subfigure}
		\caption{The posterior model probability of  the data generating model  obtained for observations generated from  (a) SIR model and (b) SEIR model according to optimal designs and an equally spaced designs.} 
		\label{fig:pmps_SL_LA_Ex2SIR_SEIR} 
	\end{figure}

	As in Example 1, the optimal designs found using the SIGT utility were assessed in terms of model discrimination and parameter estimation. 
	For each case, posterior inference was undertaken using the synthetic likelihood approach described in Section \ref{s:infFramework}.
	The posterior model probabilities of the data generating model were determined for each optimal  and equally spaced design.
	As shown in Figure \ref{fig:pmps_SL_LA_Ex2SIR_SEIR}, designs found using the SIGM utility perform well across both SIR and SEIR models.
	When the SIR model is the data generating model, SIGP designs yield less informative data sets for model discrimination while both SIGT designs and equally spaced designs perform  equally well.
	For the SEIR model, a clear difference in discrimination ability of designs is not visible.

\begin{figure}[H]
		\centering
		\begin{subfigure}[b]{0.49\textwidth}
			\centering
			\includegraphics[width=3.4in,width=3in]{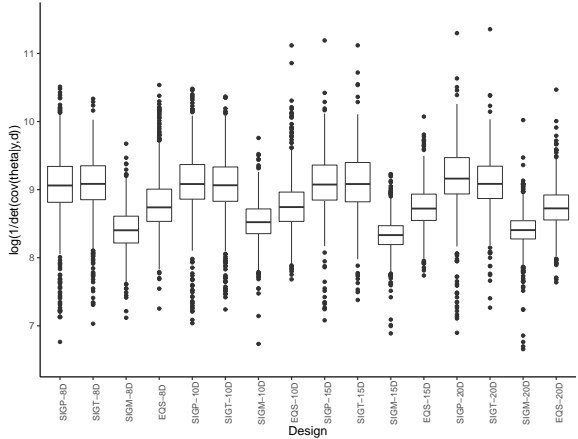}
			\caption{SIR model}
			\label{fig_Ex2SIR_PE}
		\end{subfigure}
		\hfill
		\begin{subfigure}[b]{0.49\textwidth}
			\centering
			\includegraphics[width=3.4in,width=3in]{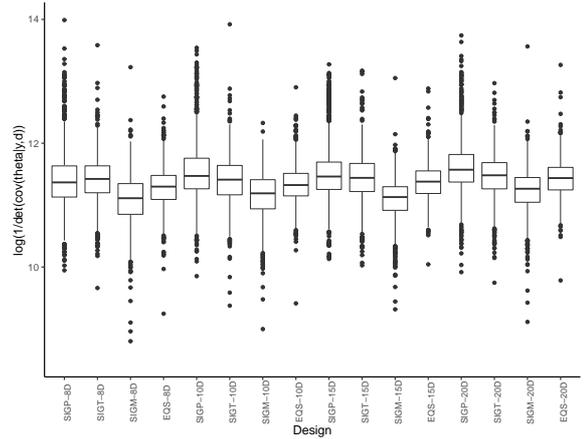}
			\caption{SEIR model}
			\label{fig_Ex2SEIR_PE}
		\end{subfigure}
		\caption{The log determinant of the inverse of posterior variance-covariance matrix of parameters of the data generating model when observations generated from  (a) SIR model and (b) SEIR model according to optimal designs and an equally spaced designs.} 
		\label{fig:LDPCV_LAIS_Ex2SIR_SEIR} 
	\end{figure}

	In order to assess the performance of the optimal designs for parameter estimation,  LIS was used as in Example 1. Figure \ref{fig:LDPCV_LAIS_Ex2SIR_SEIR} compares the log determinant of the inverse of posterior variance-covariance matrix in parameters of the data generating model based on the optimal and equally spaced designs.
	It is evident that the SIGT designs perform as well as the designs  the designs found under the SIGP utility across both models, while designs found for model discrimination (only) do not provide precise estimation of parameters, and are actually less efficient than the equally spaced designs. 
	
	\subsection{Example 3 - Design for parameter estimation of predator - prey model}
	\label{Example_3}
	
	Laboratory microcosm experiments  play a key role in developing and refining ecological theories \citep{Bonsall2005}, where single-celled organisms or insects are placed in a controlled environment to imitate complex natural environments. 
	These experiments provide many advantages over the field studies such as the ability to replicate and control environmental conditions. 
	Consequently, microcosm experiments have been used to explore ecological concepts such as  intraspecific competition \citep{Nicholson1954,Hassell1976} and
	predator and prey interaction \citep{Luckinbill1973,Lawler1993,Balciunas1995}.

	\cite{Luckinbill1973} conducted a series of experiments to investigate interactions between \textit{Pciramecium aurelia} (prey) and \textit{Didinium nasutum} (predator).
	In this example, we consider the Luckinbill's experiment as a motivating application, and find optimal sampling times to obtain data for estimating the parameters of the modified Lotka-Volterra (LV) model with logistic growth of prey.
	Let the birth rate of prey be given by  $a$ and, in the absence of predators, the prey population follows a logistic growth with a carrying capacity $K$.
	Further, the rate of predation is given by $b$ and  the death rate of predators is given by $c$.
	At time $t$, denote  the size of the prey and predators populations are $x$ and $y$, respectively, the probabilities of  possible events in the next infinitesimal time period $\Delta_t$ are given by,
	
	$$ P \big[ \, x + 1 \, , \, y \, | \, x \, , \, y \, \big] = a \, x \,\Delta_t   + o(\Delta_t), $$
	
	$$ P \big[ \, x - 1 \, , \, y \, | \, x \, , \, y \, \big] = a \bigg(1- \frac{x}{K}\bigg) \, x \, \Delta_t + o(\Delta_t), $$
	
	$$ P  \big[\, x - 1 \, , \, y + 1 \, | \, x \, , \, y \,  \big] = b\,x\,y\, \Delta_t + o(\Delta_t), $$

	$$ P  \big[\, x  \, , \, y - 1 \, | \, x \, , \, y \, \big] = c\,y \, \Delta_t + o(\Delta_t). $$
	
	
	
\begin{figure}[H]
		\begin{subfigure}[b]{1\textwidth}
			\centering
			\includegraphics[width=5in,height=2in]{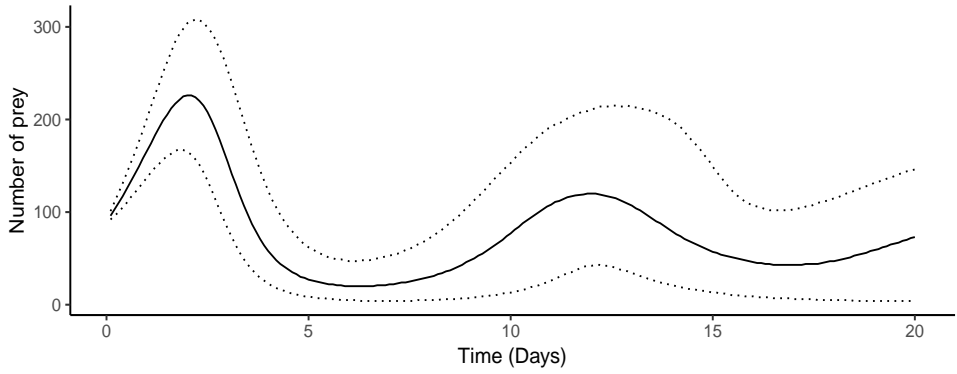}
			\caption{}
			\label{fig:P3Ex3LV2_Py}
		\end{subfigure}
		\hfill
		\begin{subfigure}[b]{1\textwidth}
			\centering
			\includegraphics[width=5in,height=2in]{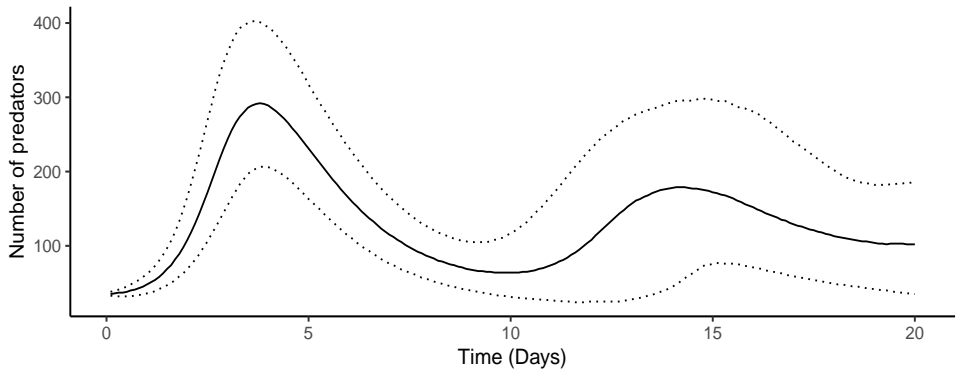}
			\caption{}
			\label{fig:P3Ex3LV2_Pd}
		\end{subfigure}
		
		\begin{subfigure}[b]{1\textwidth}
			\centering
			\includegraphics[width=6in,height=3in, keepaspectratio=false]{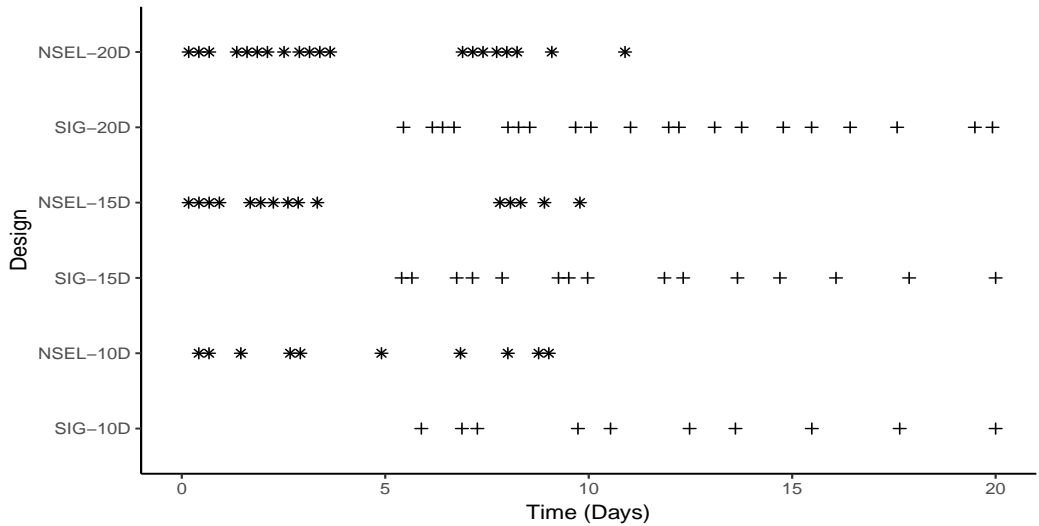}
			\caption{}
			\label{fig:OD_Ex3}
		\end{subfigure}
		\caption{Prior predictive distributions of (a) prey and (b) predators. In both figures, dotted  lines represent the $10\%$ and $90\%$ prior prediction quantiles of prey and predators. (c) Optimal designs found under SIGP ($+$) and NSEL ($*$) utilities for estimating parameters of the modified LV model. }
	\end{figure}

	Following the experimental set-up used by \cite{Luckinbill1973}, we assume that there are 90 prey and 35 predators at the beginning of the experiment. 
	To obtain oscillatory population dynamics over time, the following priors are chosen for the model parameters,
	$\log(K) \sim N(6.87,0.20^2)$ , $\log(a) \sim N(0.01,0.12^2)$ , $\log(b) \sim N(-5.03,0.12^2)$ and $\log(c) \sim N(-0.69,0.16^2)$, see Figures \ref{fig:P3Ex3LV2_Py} and \ref{fig:P3Ex3LV2_Pd} which show the distribution of prior predictive data.

	The Gillespie algorithm \citep{Gillespie1977} simulates every event that changes the state of the system.
	Compared to the epidemiological models considered in Example 1 and 2, the LV model can result in a large number of events being observed depending on the parameter values used for model simulation. 
	Consequently, for this example, the Gillespie algorithm can be computationally expensive  to simulate a large  number of times.
	Therefore, we use the Explicit tau leap (ETL) method \citep{Gillespie2001} to simulate data to estimate the mean and the covariance matrix in evaluating synthetic likelihoods.
	Here, mean, log variance and maximum of the observed counts of prey and predators according to the design ($\bmath{d}$) were considered as the summary statistics.
	
	Figure \ref{fig:OD_Ex3} shows the optimal sampling times for parameter estimation of the modified LV model found under the SIGP and NSEL utilities. 
	As can be seen, there is  overlap between the selected sampling times which maximise the SIGP and NSEL utilities.
	However, the NSEL  designs suggest to observe the process at the beginning of the experiment while SIGP designs consist of sampling times at the end of experiment. 
	Similar to previous examples, the expected utility of each optimal design was re-evaluated 100 times, and  the mean and standard deviation of these expected utility values is given in Table \ref{table:Ex3_OD_KLD_NSEL}.

	\begin{table}[h]
		\caption{ Expected utility values (standard deviation)  of optimal designs derived under the SIGP and NSEL utility functions.} \label{table:Ex3_OD_KLD_NSEL}
		{\begin{tabular*}{\columnwidth}{@{}l@{\extracolsep{1cm}}l@{\extracolsep{1cm}}r@{}}
				\hline Utility function  & Number of  design points $|\bmath{d}|$ & $U(\bmath{d}^*)$ (SD) 
				\\
				
				\hline
				\multirow{4}{*} {SIGP}  & 10 &  2.87 (0.01) \\  
				
				& 15  &     2.95 (0.02)     \\ 
				
				& 20 &   2.97 (0.02)   \\ 
				
				\hline
				\multirow{4}{*} {NSEL } & 10 &   -0.0596 (0.0004)   \\
				
				& 15 &   -0.0588 (0.0005)   \\ %
				
				& 20 &   -0.0589 (0.0005)  \\ %

				\hline
		\end{tabular*}}
		\bigskip
	\end{table}
	
\begin{figure}[H]
		\centering
		\centering
		\includegraphics[width=4.4in,width=4in]{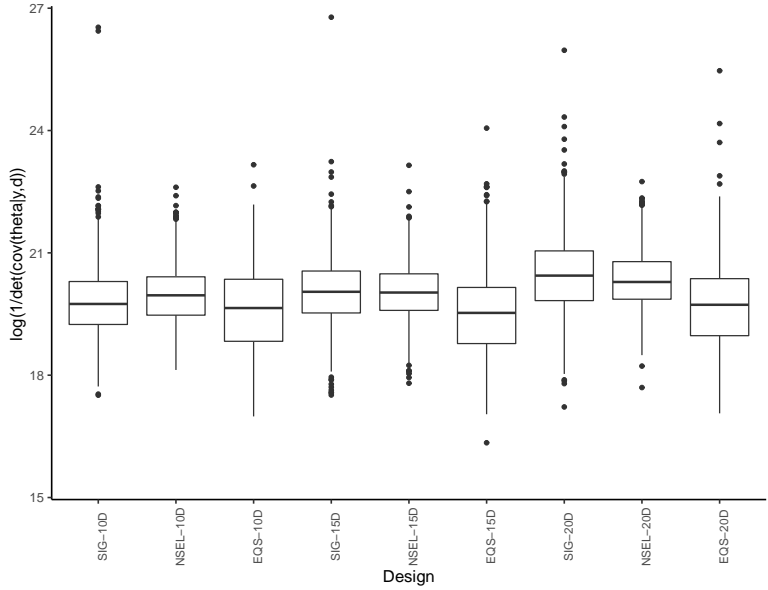}
		\caption{The log determinant of the inverse of posterior variance-covariance matrix of parameters of the data generating model when observations generated from  the modified LV model according to optimal designs and an equally spaced designs.} 
		\label{fig:LDPCV_LAIS_LV2} 
	\end{figure}

	As in the other two examples, the performance of optimal designs in estimating parameters were assessed  based on the log determinant of the inverse of posterior variance-covariance matrix of parameters.
	Figure \ref{fig:LDPCV_LAIS_LV2} compares the optimal designs found under SIGP and NSEL utilities with equally spaced designs.
	Despite noticeable differences between the SIGP and NSEL designs, they perform similarly well compared to the equally spaced designs.
	As seen in Figures \ref{fig:P3Ex3LV2_Py} and \ref{fig:P3Ex3LV2_Pd}, prior predictive distributions of prey and predators have two oscillations. 
	Thus, potentially there are two regions of the prior predictive distributions where information about parameters can be obtained.  
	Consequently, the utilities considered here select quite different regions of the design space as being informative.

	\section{Discussion}
	\label{s:discuss}
	
	In this work, we proposed a synthetic likelihood-based Laplace approximation to evaluate utility functions in designing experiments to collect a  lager number of observations in epidemiology and ecology.
	The proposed Laplace approximation requires a relativity  small number of likelihood evaluations compared to other posterior approximation methods, and this reduces the number of model simulations required for utility evaluations. 
	This approach  avoids the use of pre-simulated data sets which generally requires large storage, and may lead to sub-optimal designs as a discrete design space needs to be considered.
	Further, the computational cost in approximating the likelihoods with a large number of observations has been reduced by using summary statistics instead of full data set.  
	Consequently, our approach enables the location of high dimensional designs for models with intractable likelihoods in a continuous design space providing significant improvement on what has been proposed previously in the literature.   
	
	
	Although, the proposed approach provides an efficient approximation for a wide range of utility functions, there are a few limitations.
	First, the selection of summary statistics which are informative not only over the entire prior predictive  distribution but also across the design space appears to be a difficult task. 
	We addressed this by avoiding designs which yielded low utility values and/or poor approximations to the utility. 
	However, recently, robust methods for estimating the synthetic likelihood have been proposed, for instance, the extended empirical saddlepoint estimation \citep{Fasiolo2018}.
	Exploration of the use of such a method to improve the approximation of low utility values is a potential research avenue that could be explored into the future.

	Secondly, the use of Gillespie algorithm \citep{Gillespie1977} to simulate data can be prohibitively expensive to use in our approach, for example, see Example 3. 
	Thus, computationally less expensive methods to simulate data  such as ETL method \citep{Gillespie2001} may need to be developed in future research. 
	The exploitation of parallel computation available in Graphical processing units (GPUs) could also be useful   for alleviating some of the computational burden when finding optimal designs.  We plan to explore this into the future.

\section*{Acknowledgements}

M.B. Dehideniya was supported by a QUT Postgraduate Research Award (QUTPRA)
scholarship which is provided by QUT. Computational resources and services used in this work were provided
by the HPC and Research Support Group, QUT, Brisbane, Australia. 

\bibliographystyle{abbrvnat}
\bibliography{DualExp_TE_SL_LA}

\newpage
\begin{appendices}

\section{Informativeness of summary statistics}
\label{ISS_DMSI}
\subsection{Death model}
\begin{figure}[ht]
	
	\centering
	\begin{subfigure}[b]{0.4\textwidth}
		\centering
		\includegraphics[width=2in,width=2in]{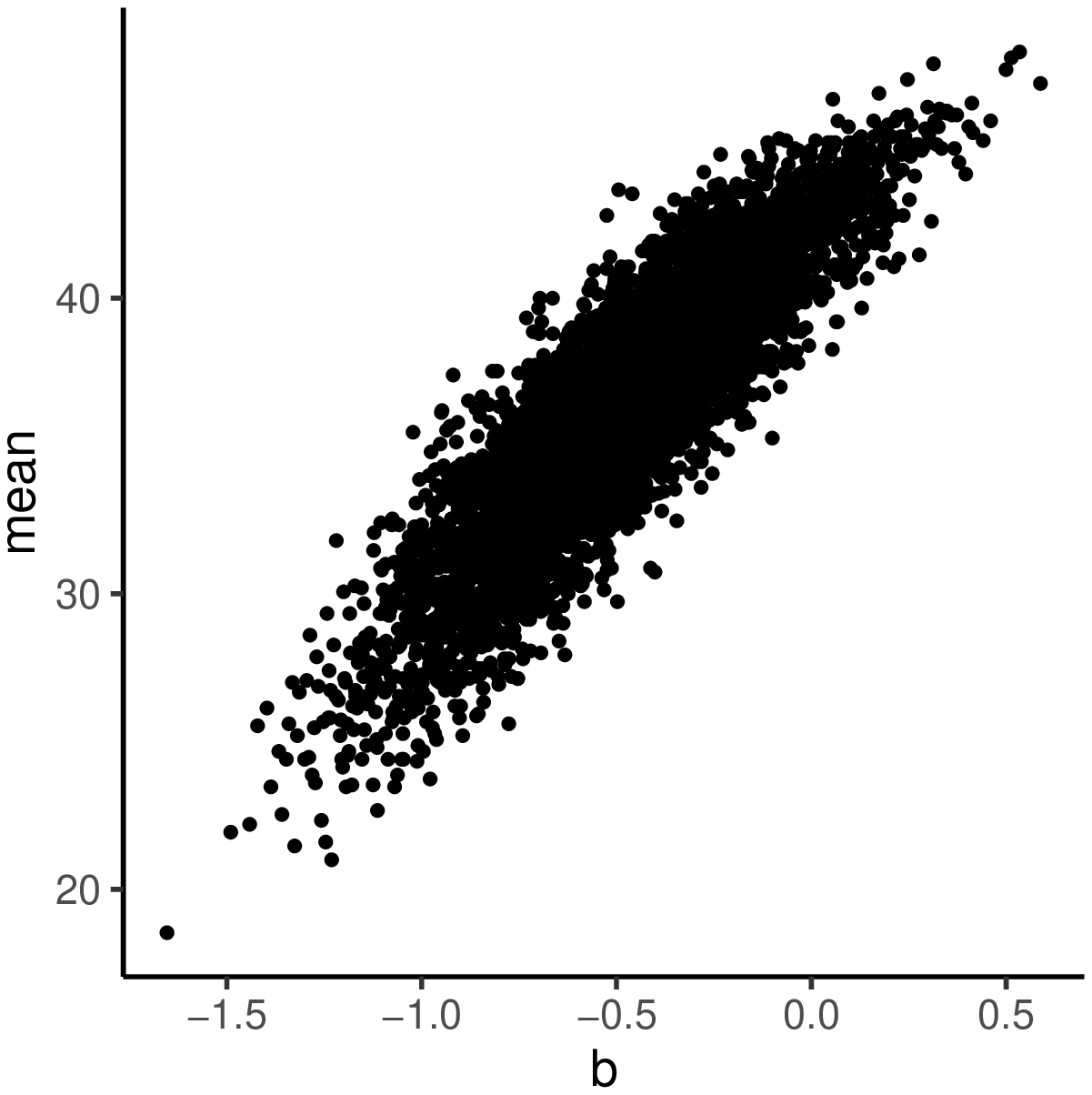}
		\caption{}
		\label{fig:bDM_Mean}
	\end{subfigure}
    \hspace{1em}
	\begin{subfigure}[b]{0.4\textwidth}
		\centering
		\includegraphics[width=2in,width=2in]{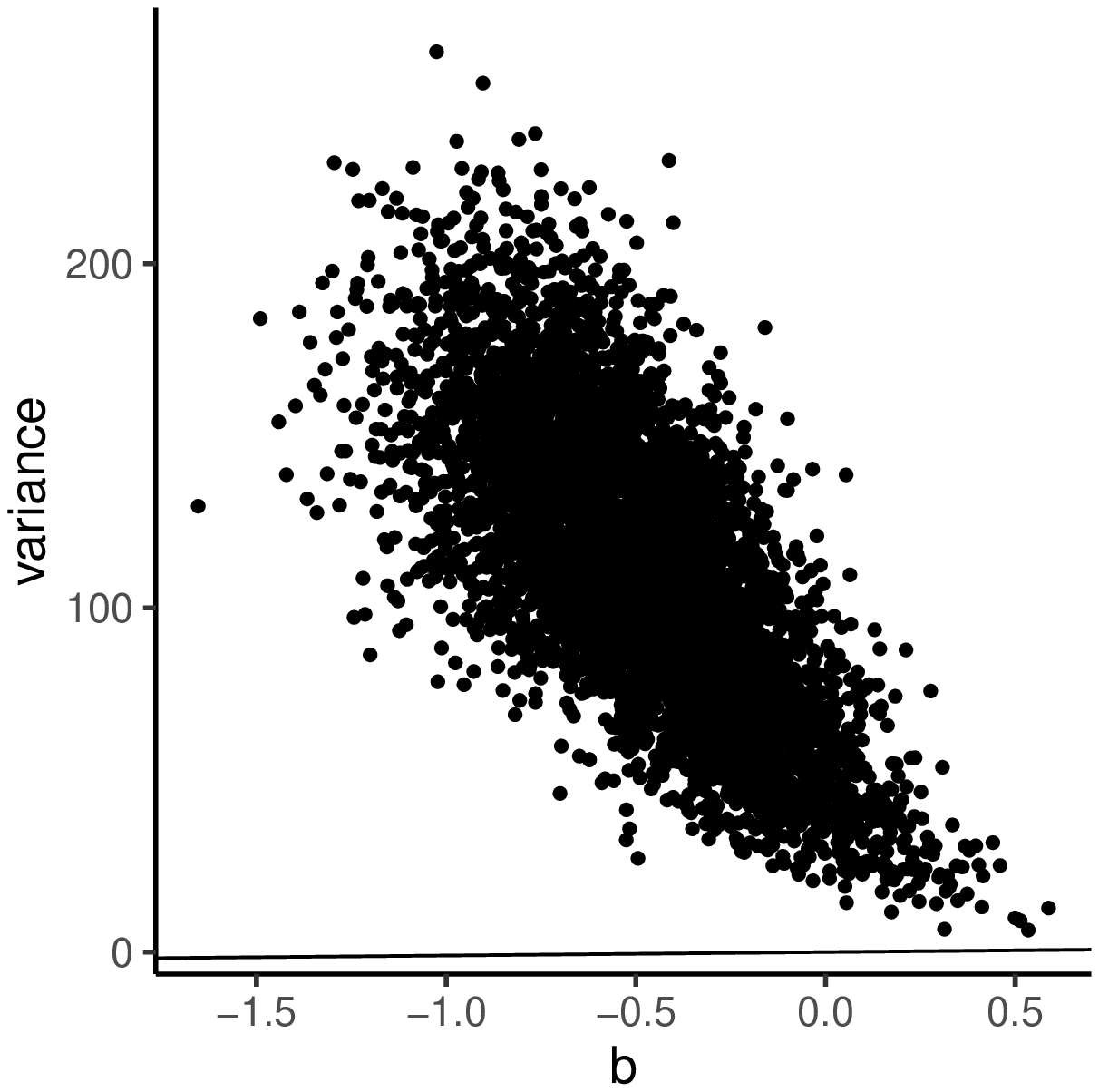}
		\caption{}
		\label{fig:bDM_Var}
	\end{subfigure}
	
	\caption{Scatter plot between model parameter $b$ and summary statistics (a) mean  and  (b) variance of observations simulated from the death model according to a random design with 15 design points.} 
	\label{fig_Ap:SS_DeathModel} 
\end{figure}

\subsection{SI model}
\begin{figure}[ht]
 
	\centering
	\begin{subfigure}[b]{0.4\textwidth}
		\centering
		\includegraphics[width=2in,width=2in]{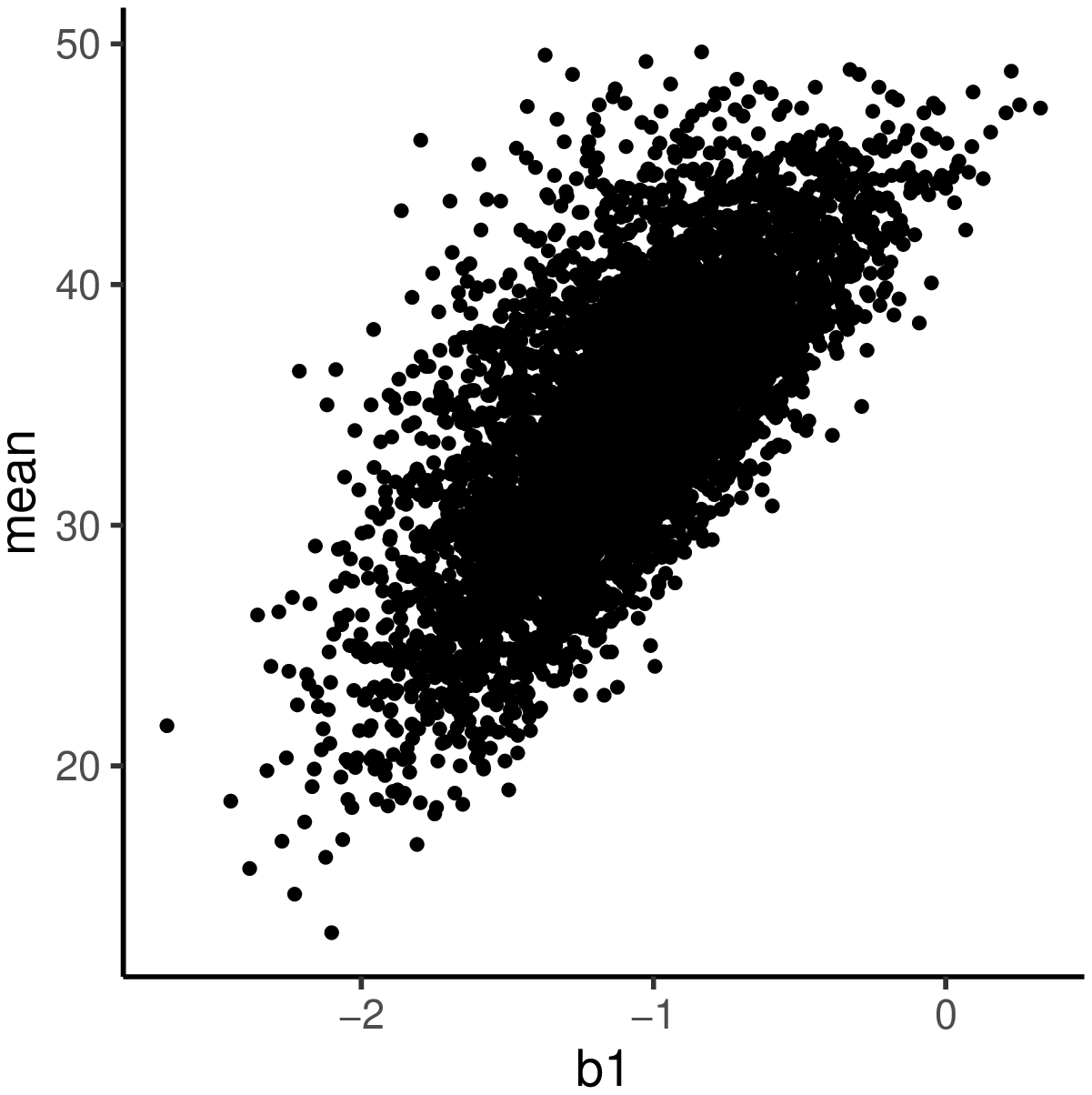}
		\caption{}
		\label{fig:b1SI_Mean}
	\end{subfigure}
	\hspace{1em}
	\begin{subfigure}[b]{0.4\textwidth}
		\centering
		\includegraphics[width=2in,width=2in]{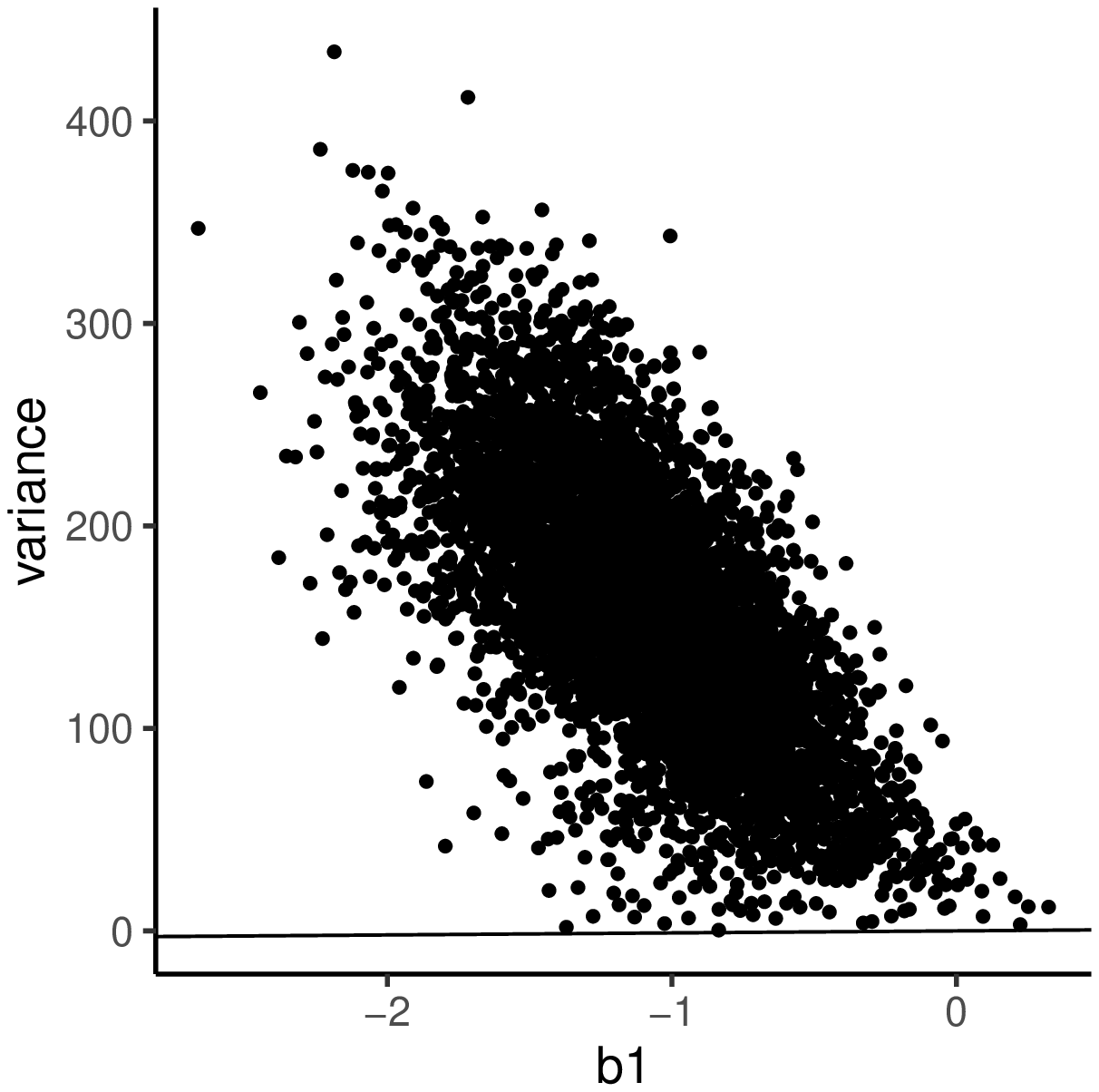}
		\caption{}
		\label{fig:b1SI_Var}
	\end{subfigure}
	
	\caption{Scatter plot between model parameter $b1$ and summary statistics (a) mean  and  (b) variance of observations simulated from the SI model according to a random design with 15 design points.} 
\label{fig_Ap:SS_SImodel_b1} 
\end{figure}

\begin{figure}[ht]
	
	\centering	
	\begin{subfigure}[b]{0.4\textwidth}
		\centering
		\includegraphics[width=2in,width=2in]{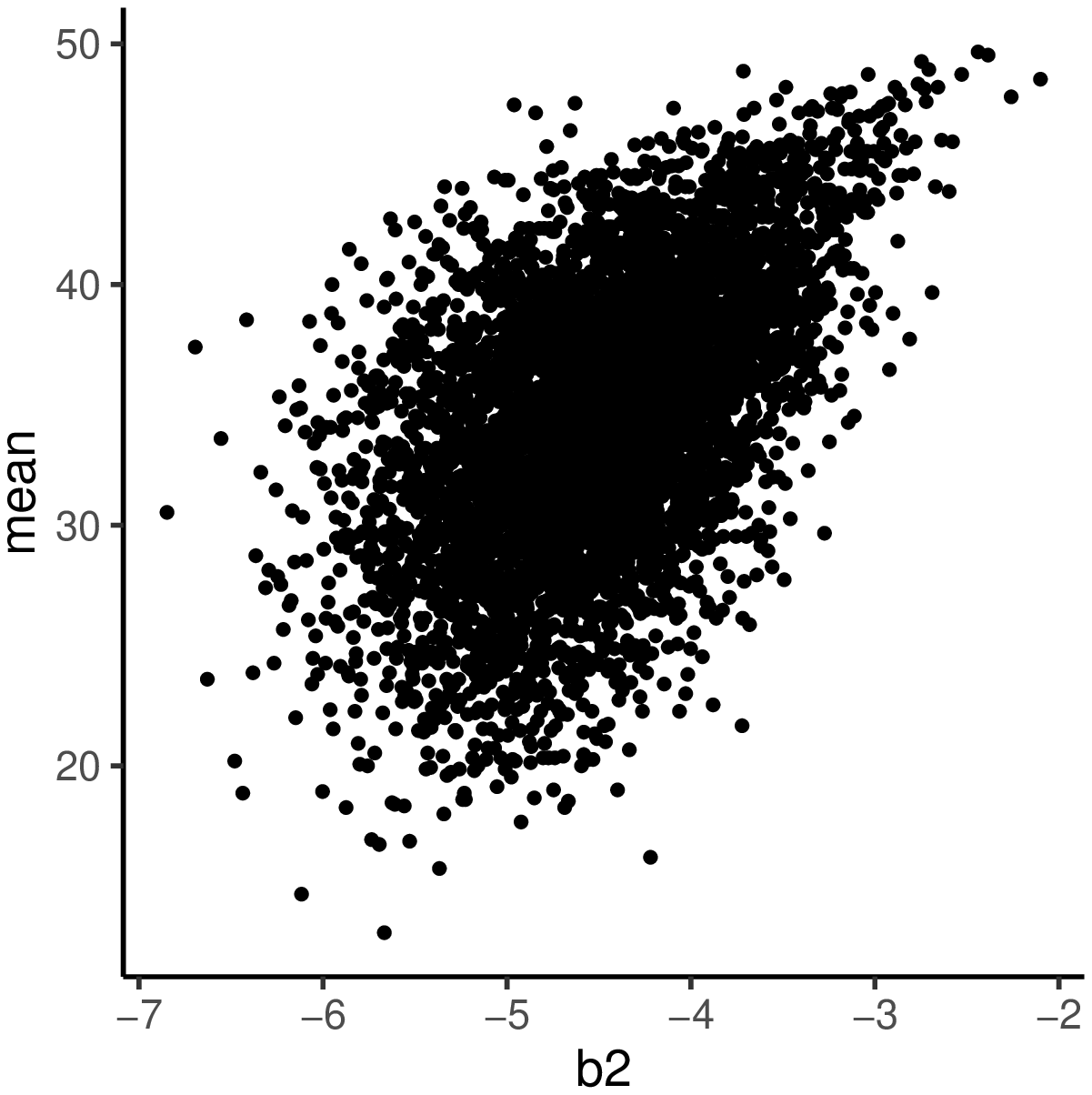}
		\caption{}
		\label{fig:b2SI_Mean}
	\end{subfigure}
	\hspace{1em}
	\begin{subfigure}[b]{0.4\textwidth}
		\centering
		\includegraphics[width=2in,width=2in]{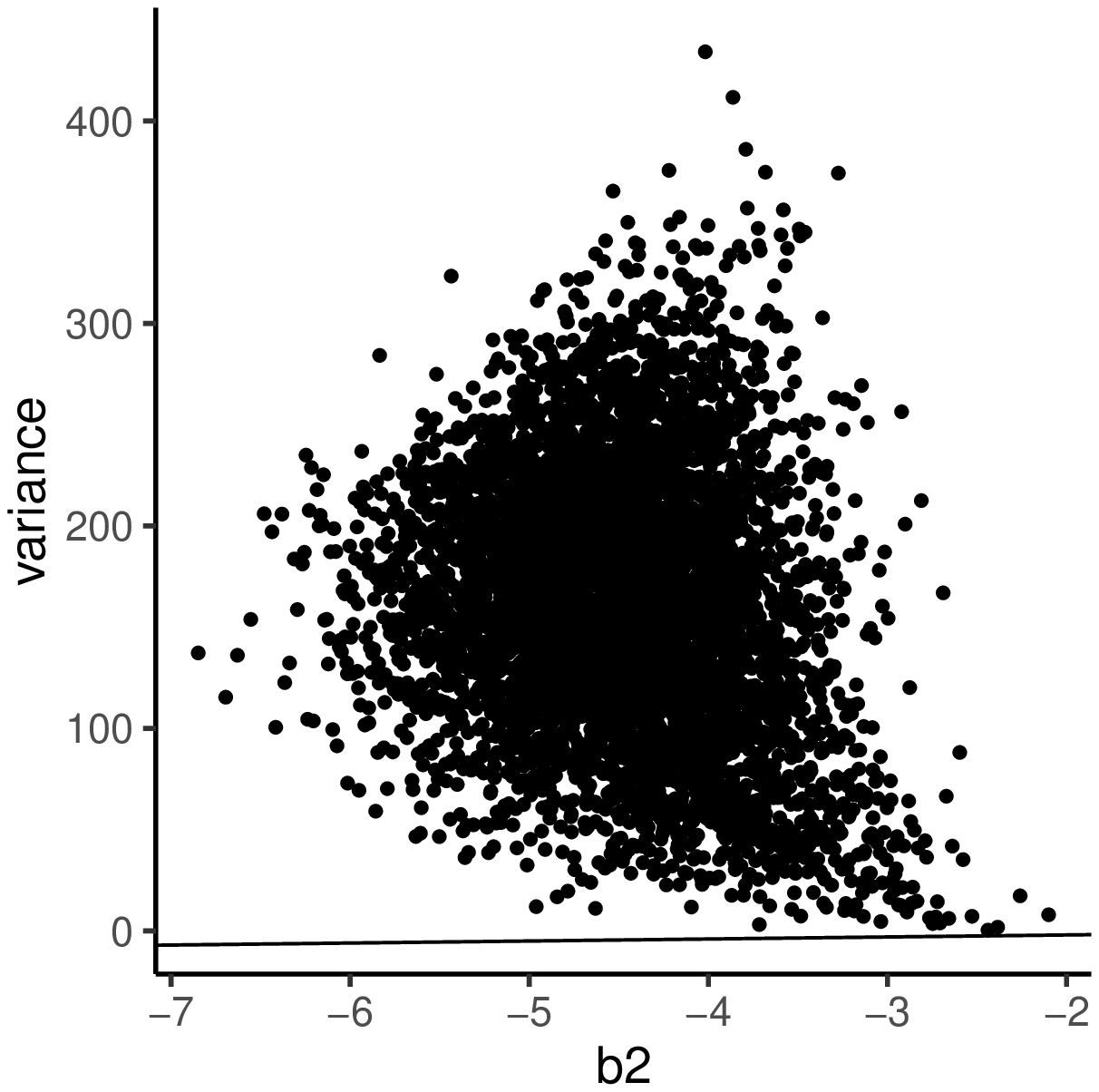}
		\caption{}
		\label{fig:b2SI_Var}
	\end{subfigure}
	\caption{Scatter plot between model parameter $b2$ and summary statistics (a) mean  and  (b) variance of observations simulated from the SI model according to a random design with 15 design points.} 
	\label{fig_Ap:SS_SImodel_b2} 
\end{figure}

\end{appendices}

\end{document}